\documentclass{article}

            \newcommand{\ad}{a^{\dag}}
            \newcommand{\cd}{c^{\dag}}
            \newcommand{\p}{\prime}

            \renewcommand{\sp}{s^{\p}}
            \newcommand{\spp}{s^{\p\p}}
            
            \newcommand{\kp}{k^{\p}}
            \newcommand{\kpp}{k^{\p\p}}
            \newcommand{\lp}{l^{\p}}
            \newcommand{\up}{u^{\p}}
            
            \renewcommand{\a}{\alpha}
            \renewcommand{\b}{\beta}
            \newcommand{\g}{\gamma}
            \newcommand{\gp}{\gamma^{\p}}
            \newcommand{\gpp}{\gamma^{\p\p}}
            
            \usepackage{graphics}
            \begin{document}
           \title{Space of Quantum Theory Representations of Natural
           Numbers, Integers, and Rational Numbers}
          \author{Paul Benioff \\
            Physics Division, Argonne National
           Laboratory,\\
           Argonne, IL 60439, USA \\
           email: pbenioff@anl.gov}
           \maketitle

           \begin{abstract}
           This paper extends earlier work on
           quantum theory representations of natural numbers $N$,
           integers $I$, and rational numbers $Ra$ to describe a
           space of these representations and transformations on
           the space. The space is parameterized by 4-tuple
           points in a parameter set. Each  point, $(k,m,h,g)$,
           labels a specific representation of $X=N,\; I,\; Ra$ as
           a Fock space $\mathcal{F}^{X}_{k,m,h}$ of states of finite
           length strings of qukits $q$ and a basis set
           $\mathcal{B}_{k,m,h,g}$ of states of $q$ strings, denoted
           together as $\mathcal{FB}^{X}_{k,m,h,g}.$ The pair $(m,h)$
           locates the $q$ strings in a square integer lattice $I\times I.$
           $k$ denotes the qukit base, and $g$ fixes the gauge or basis
           for the states of each $q.$ Basic arithmetic relations and operations
           are described for the states in $\mathcal{FB}^{X}_{k,m,h,g}.$
           Maps $(k,m,h,g)\rightarrow (k',m',h',g')$ induce transformations
           $\mathcal{FB}^{X}_{k,m,h,g}\rightarrow\mathcal{FB}^{X}_{k',m',h',g'}$
           on the representation space. There are two shifts, a base change
           operator $W_{k',k}$, and a gauge transformation function $U_{k}$ where
           $U_{k}(j,h)$ is an element of the unitary group $U(k)$.
           Invariance of the axioms and theorems for $N,\; I,$ and $Ra$
           under any transformation is discussed. It is seen that
           the properties of $W_{k',k}$ depend on the prime factors of $k'$ and
           $k$. This suggests that one consider prime number $q's,$ $q_{2},\;
           q_{3},\; q_{5},$ etc. as elementary and the base $k$ $q's$  as
           composites of the prime number $q's$.
           \end{abstract}

           \section{Introduction}

            It is quite evident that numbers play a fundamental role
            both in experimental and theoretical physics and in much
            of mathematics. There are several different types of numbers
            all of which are important.  Natural numbers are the most
           basic as they are used for counting.  Integers extend
           natural numbers to include negative whole numbers.
           Rational numbers are essential to both theory and experiment
           as they are used in most theoretical computations. They
           also appear as the output of experiments. Real and complex
           numbers are the basis of all physical theories and of many
           mathematical theories. Also all theoretical
           predictions in physics can be cast in the form of real
           number solutions to equations.

           By themselves these reasons provide support for
           investigation of the properties of numbers and
           their representations and their relation to physics.
           However, there is also an underlying motivation for this
           and related work. This is the need to understand why
           mathematics is so effective and relevant to physics.

           This problem, which was expressed by Wigner
           in 1960 \cite{Wigner}  and discussed by others
           \cite{Hamming,Fefferman}, is particularly acute if one
           accepts the widely held Platonic view that mathematical objects
           have some type of ideal, abstract existence with
           properties that are true or false in some ideal sense
           \cite{Fraenkel,Davies}. If physical existence of systems
           refers to  systems that both exist in and
           determine the properties of space time, then there is no
           reason why mathematics and physics should be related at all.
           Yet it is obvious that they are very closely related.

            There are several different approaches to understanding this
          relationship \cite{Tegmark1}-\cite{Weinberg}. The approach
          underlying this paper  is to work towards a coherent theory of
          physics and mathematics together \cite{BenTCTPM}. Such a theory,
          by treating both physics and mathematics together in one
          theory, would be expected to describe both physical and mathematical
          systems and how they are related, possibly as complementary aspects
          of a more basic type of system. It may also help to answer some of
          the basic outstanding questions in physics.

           The method used here is to study properties of numbers
           with an emphasis on some concepts that are important to physics.
           Numbers are chosen because they are so
           basic to physics, as measurement outputs, as theoretical
           predictions and as computer outputs. The representations used
           here, as states of \emph{single}, finite length strings of qukits,
           are based on the observations that all physical representations
           of numbers are by single strings of
           digits and that quantum theory is the basic theory of all
           physical systems. In addition, the use of quantum rather than
           classical representations brings both the treatment of physical
           systems and numbers into the same general theory. The use
           of the same basic theory to describe both physical systems
           and numbers as mathematical systems should help in bringing
           together descriptions of physical and mathematical systems

           The use of quantum theory to study representations
           of numbers and other mathematical systems is not new
           \cite{Litvinov} -\cite{Krol}. Of particular note is
           work on quantum set theory represented as an orthomodular
           lattice valued set theory \cite{Takeuti}-\cite{Krol}. In
           this work natural numbers, integers, and rational numbers
           have  representations that are either similar to the usual
           ones in mathematical analysis \cite{Takeuti}-\cite{Titani}
           or are based on a categorical approach \cite{Schlesinger,Krol}.
           However the work in these references differs from the approach
           taken here in that numbers are represented here as states
           of finite qukit strings.

           This work extends other work on quantum representations
           of numbers for a binary base \cite{BenRNQM}, including
           those studied in quantum computing \cite{Nielsen},  to
           descriptions for all bases $k\geq 2$. The
           study is limited to quantum representations of natural
           numbers, $N$, integers, $I,$ and rational numbers $Ra.$
           Extension to quantum representations of real and complex
           numbers will be treated in future work\footnote{Earlier
           work on quantum representations of real and complex numbers have
           been limited to states based on strings of qubits
           \cite{BenRRCNQT}.}

            The description of each quantum representation
            as a space of states of finite qukit strings, and spaces
            of these representations, is based on a parameterization
            of the representations and an association
            of a specific representation to each point in the parameter set.
            The points in the set are represented by  quadruples
            $(k,(m,h),g).$ The integer pair $(m,h)$ locates the
            string on a $2$ dimensional integer lattice, $I\times I$,
            $k$ is a natural number $\geq 2$ that is the number base,
            and $g$ is a  gauge fixing function on $N\geq 2\times I\times I$.
            For each integer pair $(j,h)$ and each $k\geq 2$,
            $g(k,j,h)$ is a basis choice for a $k$ dimensional
            Hilbert space at $(j,h).$

            Transformations  $(k,(m,h),g)\rightarrow
            (k',(m^{\p},h^{\p}),g^{\p})$  in the parameter
            set induce transformations in the representation
            space. These consist of unitary translations that
            move the qukit strings on the lattice, transformations that
            change states of strings of base $k$ qukits to states of
            strings of base $k'$ qukits, and unitary gauge transformations
            for each $k$. These are maps from the lattice points to elements
            of $U_{k}.$

            An interesting result is that the axioms and theorems
            for each of the three types of numbers are invariant
            under these transformations. They represent symmetries
            of the systems.  This is the case even though the
            specific expressions of the axioms and theorems in terms
            of basic arithmetic relations and operations are
            different for different representations. This is
            like the situation in physics where the laws of
            physics are invariant under Lorentz transformations even
            though their specific expression in different reference
            frames may be different.

            Another interesting result is that qukits
            $q_{k}$ where $k$ is a prime number function as
            elementary qukits.  These are the "elementary particles"
            as far as quantum representations of numbers  are
            concerned. Qukits where $k$ is not prime can be
            considered as composites of the prime number $q_{k}$. This
            follows from the properties of the $k$ changing
            operator, especially for rational number representations.

           The plan of the paper is as follows: In the next section
           quantum representations of the natural numbers, integers,
           and rational numbers by states of strings of base $k$ qukits
           are discussed. Here $k$ is the number of internal states of each
           qukit.

           Section \ref{SQTRN} and its subsections describe the parameter set
           and the induced space of quantum theory representations of numbers.
           Properties of  transformations  in the representation space
           associated with transformations of the parameter set are discussed
           in some detail. Of special note are the dependence of the base
           change operations on the number type ($N,I,Ra$) and on the base
           values. Commutation relations for the transformations are also
           discussed as are the special properties
           of unary representations corresponding to $k=1.$

           Section \ref{SI} discusses symmetries and invariances
           associated with the transformations. The idea is that,
           since axioms and theorems hold for each representation,
           they are invariant under any transformation.
           As such they are symmetries of the space.

           Composite and elementary qukits are discussed in the next
           section. It is noted that base $k$ qukits where $k$ is a
           prime number, play the role of elementary qukits. This is
           a consequence of the properties of sets of $Ra$ represented
           by states of single strings of kits or qukits. The prime
           number qukits can be combined into composites to give $q_{k}'s$
           for any base $k$. The paper concludes with a discussion of several
           points and a summary of what was done here.

           It is to be emphasized that all numbers are represented
           here as states of \emph{single}, finite length strings of
           qukits.  This is based on the observation that almost all
           physical representations of numbers as experimental outputs
           and as used in computers are of this form.  Representations of numbers
           by pairs of states of two finite length qukit strings are not
           used here.

           \section{Quantum Representations of Natural Numbers,
           Integers, and Rational Numbers}\label{QRNRIRN}
           Here representations of $N$, $I$, and $Ra$ are given as
           states of finite strings of base $k$ qukits on an integer
           lattice $I\times I.$ The description that follows is an
           extension of earlier work \cite{BenRRCNQT,BenRCRNQM} on qubit strings
           to qukit strings for any $k\geq 2$.

           Since states for strings of different numbers of qukits are needed,
           it is useful to describe the states using strings of annihilation
           creation (AC) operators that create or annihilate qukits in different
           states at different integer pair locations. Two types of
           qukits are used,  $(a_{k})_{\a,(j,h)},(\ad_{k})_{\a,(j,h)}$ and
           $c_{\g,(m,h)}, \cd_{\g,(m,h)}.$ Here $\a =0,1,\cdots,k-1\;\g=+,-$,
           and $j,m,h$ are integers. There is no $k$ subscript on the
           $c$ operators as they are the same qubit operators for all $k$
           values. For each value of $k,$ the AC operators can satisfy
           either commutation relations or anticommutation relations
           as the base $k$ qukits can be either bosons or fermions.
           For all $k$ the type $a$ and $c$ systems are assumed to
           be distinguishable.

\begin{figure}[h]\begin{center}
           \resizebox{100pt}{100pt}{\includegraphics[300pt,290pt]
           [540pt,530pt]{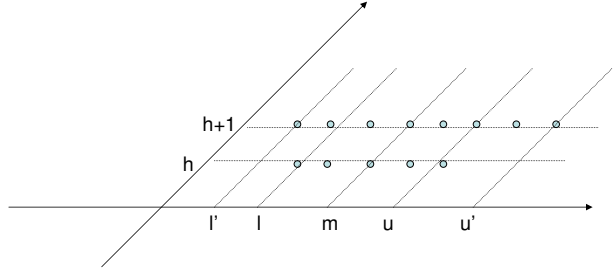}}\end{center}
           \caption{Schematic representation of two qukit strings
           on a $2$ dimensional integer lattice. The strings at
           $h$ and $h+1$ with $5$ and $8$ qukits extend from
           $j=l$ to $j=u$ and from $j=l'$ to $j=u'$ respectively.
           The site $j=m$ is occupied by both a qukit and the sign
           qubit. It is also the location of the $k-al$ point.
           The equal distance between adjacent qukits shown
           here is for illustrative purposes only. No metric
           distance between adjacent points is assumed here.}
           \label{NP1} \end{figure}

           Points on the integer lattice on which the qukits  $q_{k}$
           are located consist of pairs $(j,h)$ of integers.  Each value
           of $h$ gives the location of a string and the values of
           $j$ give the locations of the $q_{k}$ in a string.
           Figure \ref{NP1} shows schematic examples of two $q_{k}$
           strings on the lattice, one at $h$ and the other at $h+1.$

           It should be emphasized that it is not necessary here to
           assume that $I\times I$ denotes discrete points in a
           metric space. Nothing is assumed here about any metric
           spatial distance between points or even if such exists.
           The only feature of $I\times I$ used here is that it
           represents a pair of integer type orderings: Each $I$ is
           a collection of points with no least and no greatest point
           and each point has a nearest neighbor on each
           side.\footnote{Even this is more than is needed.  It is
           sufficient for one $I$ to be ordered to give an ordering
           to the $q_{k}$ in a string.  The other $I$ could be
           replaced by a denumerable set of labels that serve to
           distinguish the different strings.}

           A compact representation of numbers is used that is
           suitable for representing the three types of numbers. Here
            the integer location of the sign is also the location of
            the "k-al" point. For example $-63.71,\; 459,\; -0.0753$
            would be expressed here as $63-71,\; 459+,\;0-0753$.

            To this end let $l,m,u$ be integers where $l\leq m\leq u.$
            Define the $q_{k}$ string  state $|\g,(m,h),s,l,u\rangle_{k}$ on the
            integer pair interval $[(l,h),(u,h)]$ by \begin{equation} \label{rastrst}
            \begin{array}{l}|\g,(m,h),s,l,u\rangle_{k}=\cd_{\g,(m,h)}
            (\ad_{k})_{s(u,h),(u,h)}\cdots (\ad_{k})_{s(l,h),(l,h)}
            |0\rangle\\ \hspace{1cm}=\cd_{\g,m,h} (\ad_{k})^{s}_{[l,u]}
            |0\rangle.\end{array}\end{equation} Here $s$ is a
            $0,1,\cdots,k-1$ valued function on $[(l,h),(u,h)],$ $\g =+,-$,
            and $(m,h)$ is the location of the sign and the "k-al" point.
            The pair $(m,h)$ also serves as a useful denotation of a
            string location where the string extends from $(l,h)$ to
            $(u,h).$ In the above $\cd_{\g,m,h} (\ad_{k})^{s}_{[l,u]}$
            is a shorthand notation for the creation operator string in
            the definition.

            This representation is sufficiently broad to encompass
            all three types of numbers. It also includes qukit string
            states with leading and trailing $0s.$  For integers, $I$,
            and natural numbers, $N$, $s$ satisfies the requirement
            that $s(j,h)=0$ for all $j:l\leq j<m.$ For $N,$ $\g =+.$
             For rational number representations there are no
            restrictions on $s$ or $\g.$

            The states $|\g,(m,h),s,l,u\rangle_{k}$ and their linear
            superpositions can be regarded as elements of a Fock space,
            $\mathcal{F}^{X}_{k,(m,h)},$ that is spanned by a basis set
            $\mathcal{B}^{X}_{k,(m,h)}$ of  states
            $|\g,(m,h),s,l,u\rangle_{k}.$ Here $X=N,I,\mbox{ or }Ra,$
            $m$ and $h$ are fixed, $l$ and $u$ can vary
            with $l\leq m\leq u$, and, for $X=Ra$, $s$ is
            any $0,\cdots k-1$ valued function on the interval
            $[(l,h),(l+1,h),\cdots,(u,h)].$  For $X=I$ and $N$, $s$
            satisfies the restriction given above. Note that the
            states $|\g,(m,h),s,l,u\rangle_{k}$ are pairwise
            orthogonal, \begin{equation}\label{orthog} _{k}\langle
            \g',(m',h')s',l',u'|\g,(m,h),s,l,u\rangle_{k}=\delta_{\g',\g}
            \delta_{s',s}\delta_{(m'h'),(m,h)}\delta_{(l',u'),(l,u)}.
            \end{equation} The Fock space representation is used
            because linear superpositions of states with different
            numbers of $q_{k}$ are included.

            In the following the action of operators on pairs and
            triples, and more generally $n-tuples$ of these states
            needs to be considered.  States representing pairs or
            $n-tuples$ of $q_{k}$ strings, at different $h$ values
            can either be represented by extension of the Fock space
            representation or by considering tensor products of the
            Fock spaces $\mathcal{F}^{X}_{k,(m,h)},$.  In either
            case the basic units are the basis states given by Eq.
            \ref{rastrst} where $\cd_{\g,m,h} (\ad_{k})^{s}_{[l,u]}$
            acts like a $q_{k}$ string state creation operator.
            (This description is assumed to also include the sign
            qubit implicitly.)

            The Fock space extension for pairs of states of $q_{k}$
            strings, one at $(m',h')$ and the other at $(m,h),$ is represented
            by \begin{equation}\label{pair}
            |\gp,(m',h'),s',l',u';\g,(m,h),s,l,u\rangle_{k} =
            \cd_{m',h'} (\ad_{k})^{s'}_{[l',u']}\cd_{m,h}
            (\ad_{k})^{s}_{[l,u]}|0\rangle.\end{equation} Here the
            only restriction is that the two strings do not overlap
            for either bosons or fermions. This is accounted for by
            requiring that $h\neq h'.$ There is no restriction on
            the values of $m$ and $m'$.

            The other representation, which is the one which will be
            used here, is to represent pairs of
            $q_{k}$ string states  by $|\gp,(m',h'),s',l',u'\rangle_{k}
            |\g,(m,h),s,l,u\rangle_{k}.$ In general, $n-tuples$ have a similar
            representation. These states and their linear superpositions
            are elements of tensor products of the spaces $\mathcal{F}^{X}_{k,(m,h)}.$

            The correspondence between the two representations is
            given by \begin{equation}\label{pairprod}
            |\gp,(m',h'),s'l',u';\g,(m,h),s,l,u\rangle_{k}\Leftrightarrow
            |\gp,(m',h'),s'l',u'\rangle_{k}|\g,(m,h),s,l,u\rangle_{k}
           \end{equation} for pairs. For $n-tuples$ the
           correspondence is given by\begin{equation}\label{ntupleprod}
           \begin{array}{l}|\g_{1},(m_{1},h_{1}),s_{1},l_{1},u_{1};\cdots ;
            \g_{n},(m_{n},h_{n}),s_{n},l_{n},u_{n}\rangle_{k}\\
            \hspace{1cm}=|\g_{1},(m_{1},h_{1}),s_{1},l_{1},u_{1}\rangle_{k}\cdots
            |\g_{n},(m_{n},h_{n}),s_{n},l_{n},u_{n}\rangle_{k}.\end{array}
           \end{equation}  From now on the product state
           representation will be used.

            The basic arithmetic relations and operations consist of
            equality $=,$ and an ordering for all three number types,
            and addition $+$ and multiplication, $\times$, for the
            natural numbers.  For the integers, subtraction $-$, is
            added, and for rational numbers, division $\div,$ is added.
            The properties of these relations and operations are given
            by the three sets of axioms for natural numbers, $N$,\footnote{For
            $N$ a successor operation should also be included.  This
            operation, and its use to define the other arithmetic
            operations, is discussed in \cite{BenRNQM}} integers,
            $I$, and rational numbers, $Ra.$

            There are two ways to approach the problem of showing
            that the states $|\g,(m,h),s,l, u\rangle_{k}$ represent
            numbers.  One way is to define an operator
            $\tilde{N}_{k}$ whose eigenstates are the states
            $|\g,(m,h),s,l, u\rangle_{k}$ and whose eigenvalues are the
            $N,I,$ or $Ra$ equivalent numbers in the real number base $R$
            of the Fock spaces. One knows that the sets of
            eigenvalues of $\tilde{N}_{k}$ as  subsets of $R$
            satisfy the relevant axioms for the arithmetic
            relations and operations on the relevant subsets of $R$.
            This can be used to require that the  arithmetic relations and
            operations on the states $|\g,(m,h),s,l, u\rangle_{k}$
            be defined so that the operator $\tilde{N}_{k}$ is an
            arithmetic isomorphism.  That is, it preserves the basic
            arithmetic relations and operations. If this is true, then
            it follows that the states represent numbers as they satisfy the
            relevant axioms.

            An example of an operator  for this role is given by the
            definition of $\tilde{N}_{k}$ as the product of two
            commuting operators, a sign scale operator
            $(\tilde{N}_{k})_{ss}$, and a value operator $(\tilde{N}_{k})_{v}.$
            One has\begin{equation} \label{defN}\begin{array}{c}\tilde{N}_{k}
             =(\tilde{N}_{k})_{ss}(\tilde{N}_{k})_{v} \\ \mbox{where }
            (\tilde{N}_{k})_{ss}=\sum_{\g,m}\g k^{-m}\cd_{\g,m}c_{\g,m} \\
            \tilde{N}_{v}= \sum_{\a,j,h}\a k^{j}(\ad_{k})_{\a,(j,h)}
            (a_{k})_{\a,(j,h)}. \end{array}\end{equation} Note that,
            because of the presence of strings of leading or trailing $0s,$
            the eigenspaces of $\tilde{N}$ are infinite dimensional.
            The eigenspace for the number $0$ includes the state
            $|0+\rangle =\cd_{+,m}(\ad_{k})_{0,m}|0\rangle$ and all
            states of the form $|\g,(m,h),s,l,u\rangle_{k}$ where
            $s$ is the constant $0$ function $\bar{0}_{[l,u]}$ of
            $0s$ from $l$ to $u$.

            The other way to show that the states
            $|\g,(m,h),s,l,u\rangle_{k}$ represent numbers is to
            define basic arithmetic relations and operations in
            terms of simple operations on the states and show that
            the definitions given do satisfy the relevant axioms.
            The main advantage of this approach is that it is more
            direct in that it makes no use of elements of $R$ and their
            arithmetic properties. This more direct approach is the
            one used elsewhere \cite{BenRNQM,BenRRCNQT} and is followed here.
            However, here the definitions in terms of simple
            operations are included by reference so as not to repeat
            other work.  In any case one notes that the assertion
            that the states $|\g,(m,h),s,l,u\rangle_{k}$ represent
            numbers is relative to the definitions of the arithmetic
            relations and operations and their satisfaction of
            relevant axioms.  It has no absolute or intrinsic
            meaning.

            In the following the values of $l,u$ will often be
            suppressed in state representations. In this case the state
            $|\g,(m,h),s,l,u\rangle_{k}$ will be represented
            in a short form as $|\g,(m,h),s\rangle_{k},$ where the
            values of $l,u$ are included in the definition of $s.$

            The basic relations and operations are defined for
            each value of $k$ and for different values of $(m,h).$
            Arithmetic equality, $=_{A,k,m,h,h'},$ as a
            binary relation, is defined between states in
            $\mathcal{F}^{X}_{k,m,h}$ and  $\mathcal{F}^{X}_{k,m,h'}$
             by\begin{equation}\label{equalA}
            \begin{array}{c}|\g,(m,h),s\rangle_{k} =_{A,k,m,h,h'}|\g^{\p},(m,h'),s^{\p}
            \rangle_{k}, \\ \mbox{if $\g^{\p}=\g$ and for all $j$} \\
            \mbox{If $j$ is in  both $[l,u]$ and $[l',u'],$ then
            $s(j,h)=\sp(j,h'),$}\\ \mbox{ If $j$ is in  $[l,u]$ and not in $[l',u'],$
            then $s(j,h)=0.$}\\ \mbox{If $j$ is in  $[l',u']$ and not in $[l,u],$
            then $s'(j,h')=0.$}\end{array} \end{equation} Here $X=N,I,$ or $Ra.$
            This definition is complex because it defines equality up to
            leading and trailing $0s.$ The extension of the definition to
            different m values as in $=_{A,k,m,h,m',h'}$ is more
            complex in that the difference between $m$ and $m'$ must
            be taken into account.

            Note that quantum state equality implies arithmetic equality:
            $$ (|\g,(m,h),s\rangle_{k} =|\g^{\p},(m,h'),s^{\p}\rangle)_{k}\rightarrow
            (|\g,(m,h),s\rangle_{k} =_{A,k,m,h,h'}|\g^{\p},(m,h'),s^{\p}\rangle_{k}).$$
            However the converse implication does not hold.

         Arithmetic ordering $\leq_{A,k,m,h,h'}$ on $N$, and on
         positive $I$ and $Ra$ states is defined by
         \begin{equation}\label{deforderA}\begin{array}{l}|+,(m,h),s\rangle_{k}
         \leq_{A,k,m,h,h'}|+,(m,h'),s^{\p}\rangle_{k}\\ \hspace{0.5cm} \leftrightarrow
         \left(\begin{array}{l}|+,(m,h),s\rangle_{k}
         =_{A,k,m,h,h'}|+,(m,h'),s^{\p}\rangle_{k}\mbox{ or }\\ |+,(m,h),s\rangle_{k}
         <_{A,k,m,h,h'}|+,(m,h'),s^{\p}\rangle_{k}\end{array}\right.\end{array}
         \end{equation} where\begin{equation}\label{deforderA1}
         \begin{array}{l}|+,(m,h),s\rangle_{k}
         <_{A,k,m,h,h'}|+,(m,h'),s^{\p}\rangle_{k}
         \mbox{ if}\\ \hspace{0.5cm}\mbox{for some $j$ in both $[l,u]$ and
         $[l',u'],$}\\ \hspace{1cm}\mbox{$s(j,h)<\sp(j,h')$ and $s_{[(j+1,h),(u,h)]}=
         \sp_{[(j+1,h'),(\up,h')]}$}\\ \hspace{1.5cm}
         \mbox{ up to leading $0s.$}\end{array} \end{equation}
         The extension to zero and negative $I$ and
         $Ra$ states is given by\begin{equation}\label{negordr}
         \begin{array}{c}|+,(m,h),0\rangle_{k}
         \leq_{A,k,m,h,h'}|+,(m,h'),s'\rangle_{k}
         \mbox{ for all  $s'$}\\ |+,(m,h),s\rangle_{k}
         \leq_{A,k,m,h,h'}|+,(m,h'),s^{\p}\rangle_{k}\\
         \rightarrow |-,(m,h'),s^{\p} \rangle_{k}\leq_{A,k,mh.h'}
         |-,(m,h),s\rangle_{k}.\end{array}\end{equation}

          The definitions of the arithmetic relations, $=_{A,k,m,h,h'},
          \leq_{A,k,m,h,h'}$ are given for specific values of
          $m,h,$ and $h'$.  These restrictions can be removed by
          extending the definitions to apply also to arbitrary
          values of $(m,h)$ and $(m',h').$  The only restriction is that any
          pair of strings being compared have no overlap. This is
          the case if and only if $h\neq h'.$ This is the case
          because the values of $h$ distinguish the different
          strings whereas the values of $m$ locate the sign and
          $"k=al"$ point in a string.

          In the following it is often useful to define
          arithmetic relations that are equivalent to sums over all
          pairs of $h,\neq h'.$ That is\begin{equation}\label{globaleqlt}
          \begin{array}{l}=_{A,k,m}\leftrightarrow \exists {h,h'}
          =_{A,k,m,h,h'}\\ \leq_{A,k,m}\leftrightarrow \exists {h,h'}
          \leq_{A,k,m,h,h'}.\end{array}\end{equation}

         Projection operators $\tilde{P}_{=_{A,k,m}}$ and $\tilde{P}_{
         \leq_{A,k,m}}$ can be associated with these relations. The
         action of $\tilde{P}_{=_{A,k,m}}$ on pairs of basis states for
         arbitrary $h,h'\neq h$ is given
         by\begin{equation}\label{Peqbin}\begin{array}{l}
         \tilde{P}_{=_{A,k,m}}|\g,(m,h),s''\rangle_{k}|\gp,(m,h'),\sp\rangle_{k}
         =\\ \left\{\begin{array}{l}|\g,(m,h),s''\rangle_{k}|\gp,(m,h'),
         \sp\rangle_{k}\\ \hspace{1cm}\mbox{ if } |\g,(m,h),s''\rangle_{k}=_{A,k,m}
         |\gp,(m,h'),\sp\rangle_{k}, \\ 0 \mbox{ if }|\g,(m,h),s''\rangle_{k}\neq_{A,k,m}
         |\gp,(m,h'),\sp\rangle_{k}.\end{array}\right. \end{array}\end{equation}
         $\tilde{P}_{=_{A,k,m}}$ can also be written as\begin{equation}\label{PeqAkmg}
         \tilde{P}_{=_{A,k,m}}=\sum_{\g,s^{diff}}\sum_{h'>h}\tilde{P}_{\g,[s],k,h}
         \tilde{P}_{\gamma,[s],k,h'}\end{equation}
         where\begin{equation}\label{Pssp}\tilde{P}_{\g,[s],k,h}=\sum_{s'\sim_{0}s}
         \tilde{P}_{|\g,(m,h),s',l,u\rangle_{k}}.\end{equation} In
         terms of A-C operators $\tilde{P}_{|\g,(m,h),s',l,u\rangle_{k}}$
         can be expressed as \begin{equation}\label{ACeq}
         \tilde{P}_{|\g,(m,h),s',l,u\rangle_{k}}=\cd_{\g,m,h}
         (\ad_{k})^{s}_{[l,u]} (a_{k})^{s}_{[l,u]}c_{\g,m,h}.\end{equation}
         Here $\cd_{\g,m,h} (\ad_{k})^{s}_{[l,u]}$ is given by Eq.
         \ref{rastrst}. In these equations the sum over $s^{diff}$
         is restricted to those $s$ that have no leading or trailing
         $0s$. The sum over $s'\sim_{0}s$ is over all $s'$
         that differ from $s$ only by leading or trailing $0s.$

         This definition can be applied to linear superposition
         states.  The probability that $\psi=_{A,k,m}\phi$ where
         $\psi=\sum_{\g,s}c_{\g,s}|\g,(m,h),s\rangle_{k}$ and
         $\phi=\sum_{\g,s}d_{\g,s}|\g,(m,h'),s\rangle_{k}$ is
         given by \begin{equation}\label{psiphieq}\langle
         \psi\phi|\tilde{P}_{=_{A,k,m}}|\psi\phi\rangle
         =\sum_{\g,s^{diff}}\sum_{s'\sim_{0}s}\sum_{s''\sim_{0}s}
         |c_{\g,s'}|^{2}|d_{\g,s''}|^{2}\end{equation}

         As a simple example let $\psi=1/\sqrt{2}(|22+\rangle +
         |022+\rangle)$ and $\phi =1/\sqrt{2}(|22+0\rangle+|121+
         \rangle).$ The probability that these two states are
         arithmetically equal is $1/2$ even though they are
         quantum mechanically orthogonal.

         An equation similar to Eq.
         \ref{Peqbin} holds for $\tilde{P}_{\leq_{A,k,m}}$ where
         \begin{equation}\label{Pleqbin}\tilde{P}_{\leq_{A,k,m}}=
         \sum_{\g,s,\gp,\sp:\; \g,s\leq\gp,\sp}\tilde{P}_{[\g,s]}
         \times\tilde{P}_{[\gp,\sp]}.\end{equation}
         Here $\g,s\leq\gp,\sp$ is defined by Eqs.
         \ref{deforderA}-\ref{negordr}.

          The basic arithmetic operations are $+,$ $-,$ and $\times.$
          Division will be considered later. For each $k$ and $m,$
          unitary operators for $+,$ $-,$ and $\times,$
          are represented by $\tilde{+}_{A,k,m},\;\tilde{-}_{A,k,m},$
          and $\tilde{\times}_{A,k,m}.$ These operators act on pairs
          of $q_{k}$ string states as input. Output consists of the
          pair of input states and a result string states. To
          express this in more detail, let $\tilde{O}_{A,k,m}$
          represent any of the three operations, ($O=+,-,$ or
          $O=\times.$) Then
          \begin{equation}\label{arithops}\begin{array}{l}
         \tilde{O}_{A,k,m}|\g,(m,h), s\rangle_{k}|\gp,(m,h'),\sp\rangle_{k}
         \\ \hspace{0.5cm}=|\g,(m,h),s\rangle_{k}|\gp,(m,h'),\sp
         \rangle_{k}|\g^{\p\p},(m,h''), s^{\p\p}\rangle_{k,O}\end{array}
         \end{equation} The preservation of the input
         states is sufficient to ensure that the operators are
         unitary. The values of $h,h',h''$ are arbitrary except that
         they are all different.

         In these equations the states $|\gpp,(m,h''),\spp\rangle_{k}$ with
         subscripts $O=+,-,\times$ give the results of the arithmetic
         operations.  It is often useful to write them as
         \begin{equation}\label{addnplA}\begin{array}{l}
         |\g^{\p\p},(m,h''),s^{\p\p}\rangle_{k,+}= |(m,h''),(\gp,
         \sp+_{A}\g,s)\rangle_{k},\\ |\g^{\p\p},(m,h''),s^{\p\p}
         \rangle_{k,-}= |(m,h''),(\gp,\sp -_{A}\g, s)\rangle_{k},
         \\ |\g^{\p\p},(m,h''),s^{\p\p}\rangle_{k,\times}=
         |(m,h),(\gp,\sp \times_{A}\g,s)\rangle_{k}.\end{array}
         \end{equation} The subscript $A$ on these operations
         distinguishes them as arithmetic operations. They are different
         from the quantum operations of linear superposition, $+,-$
         and product, $\times$ with no subscripts.

         The action of these linear operators on general
         states $\phi =\sum_{\g,s}c_{\g,s}|\g,(m,h),s\rangle_{k}$,
         $\psi=\sum_{\gp,\sp}d_{\gp,\sp}|\gp,(m,h'),\sp\rangle_{k}$ is given by
         \begin{equation}\label{oplin} \begin{array}{l}\tilde{O}_{A,k,m}\phi\psi
         =\sum_{\g,s}\sum_{\gp,\sp}c_{\g,s}d_{\gp,\sp}|\g,(,m,h),s\rangle_{k}\\
         \hspace{0.5cm}\times|\gp,(m,h'),\sp\rangle_{k}|(,m,h''),
         (\g,s O_{A}\gp,\sp)\rangle_{k}.\end{array}\end{equation}

         For linear superposition states arithmetic properties, including
         those expressed by the axioms and theorems, are true with some
         probability between $0$ and $1$.  For example the probability that
         $\phi=_{A,k,m}\psi$ is true is given by $(\phi\psi|\tilde{P}_{=_{A,k,m}}
         |\phi\psi)$ with $\tilde{P}_{=_{A,k,m}}$ given by Eq. \ref{Peqbin}.
         A similar expression holds with $\leq_{A,k,m}$ replacing $=_{A,k,m}.$

         Properties involving the arithmetic operations have complex
         expressions because the operations induce entanglement. For example,
         the probability that the axiom, $x+y=y+x,$ that expresses the
         commutativity of addition, is true is obtained as follows:  One needs to
         compare the result of the addition $\tilde{+}_{A,k,m}\phi\psi$
         with the result of the addition $\tilde{+}_{A,k,m}\psi\phi$. These
         two results are obtained by carrying out the additions and
         taking the traces over both input components.  For $\phi$ and
         $\psi$ the results are given by two density operators($m$ and
         $h$ are suppressed in the states):\begin{equation}\label{rhocommeq}
         \begin{array}{l}\rho_{\phi+_{A,k}\psi}=Tr_{|\g,s\rangle_{k},
         |\gp,\sp\rangle_{k}}(\tilde{+}_{A,k,m}\phi, \psi)\\ =
         \sum_{\g,s}\sum_{\gp,\sp}|c_{\g,s}|^{2}|d_{\gp,\sp}|^{2}
         |\g,s+_{A}\gp,\sp\rangle_{k}\langle\g,s+_{A}\gp,\sp|
         \\ \rho_{\psi+_{A,k}\phi}=Tr_{|\gp,\sp\rangle_{k},
         |\g,s\rangle_{k}}(\tilde{+}_{A,k,m}\psi,
         \phi)= \\ \sum_{\g,s}\sum_{\gp,\sp}|d_{\g',s'}|^{2}|c_{\g,s}|^{2}
         |\gp,\sp+_{A,k}\g,s\rangle_{k}\langle\g,s+_{A,k}\gp,
         \sp|.\end{array}\end{equation} For these two density
         operators the probability that\begin{equation}\label{Treqcomm}
         \rho_{\phi+_{A,k}\psi}=_{A,k}\rho_{\psi+_{A,k}\phi}\end{equation}
         is true is given by\begin{equation}\label{eqplusphps}
         Tr(\tilde{P}_{=_{A,k,m}}
         \rho_{\phi+_{A,k}\psi}\times\rho_{\psi+_{A,k}\phi})\end{equation}
         with $\tilde{P}_{=_{A,k,m}}$ given by Eq. \ref{Peqbin}.
         The trace is taken over all remaining variables.  Note that
         in evaluating this by carrying out the sums involved, one
         uses the commutativity of addition for basis states,
         $|\g,s+_{A,k}\gp,\sp\rangle_{k}=_{A,k,m}
         |\gp,\sp+_{A,k}\g,s\rangle_{k}.$

         The division operator $\tilde{\div}_{A,k,m},$ which
         applies to  the rational numbers only, satisfies
         \begin{equation}\label{defdiv}
         \begin{array}{l}\tilde{\div}_{A,k,m}|\g,(m,h),s\rangle_{k}
         |\gp,(m,h'),s'\rangle_{k}\\ \hspace{0.5cm} =|\g,(m,h),s\rangle_{k}
         |\gp,(m,h'),s'\rangle_{k} |(m,h''),(\g,s\div_{A}\gp,s')\rangle_{k'}\\
         \hspace{1.2cm} \mbox{where }k'=k'(k,|\gp,(m,h'),s'\rangle).
         \end{array}\end{equation} This definition has the advantage
         that it satisfies the usual axioms of division, such as
         closure under division.  However it has the problem
         that the quotient state
         $|(m,h''),(\gp,\sp\div_{A}\g,s)\rangle_{\kp}$ resulting from dividing
         $|\gp,\sp\rangle_{k}$ by $|\g,s\rangle_{k}$ is, in many cases, not a
         base $k$ state. Instead it is a base $\kp$ state where
         $\kp$ is a function of both $k$ and the state
         $|\g,(m,h),s\rangle_{k}.$ In addition, this representation
         of division, applied to linear superposition states as
         divisors, results in base entanglement.

         Examples are the best way to see this. The
         base $10$ inverse of $13$ is given by,\begin{equation}
         \tilde{\div}\label{ex1div}|1+0\rangle_{10}
         |13+\rangle_{10}=|1+0\rangle_{10}|13+\rangle_{10}|0+(01)
         \rangle_{13}\end{equation} As shown the result corresponds to the
         base $13$ number $0.1$ or as $0.(01)$. The second example
         corresponds to the base $10$ division $1740\div 13$
         \begin{equation}\label{ex2div}\begin{array}{l}
         \tilde{\div}|(1)(7)(4)(0)+\rangle_{10}
          |(1)(3)+\rangle_{10}\\ \hspace{1cm}=|(1)(7)(4)(0)+\rangle_{10}|(1)(3)
         +\rangle_{10}|(10)(03)+(11)\rangle_{13}.\end{array}\end{equation}
         The base $13$ answer corresponds to $0.1\times 1740$
         converted to base $13$ which is $(10)(03).(11)$.

         This use of decimal numbers in parentheses is a convenient way to
         denote base $k$ digits.  One needs $k$ distinct digits
         $0,1,\cdots,k-1$. The idea is to let numbers in parentheses
         denote the digits.  Thus $(31)$ denotes the $32nd$ digit for
         any base $k\geq 32,$ etc.

         A simple example to show both base and state entanglement
         is the division  of $|(1)(7)(4)(0)+\rangle_{10}$ by the state
         $1/\sqrt{2}(|(1)(3)+\rangle_{10}+|(2)(1)+\rangle_{10}).$
         The result, given by\begin{equation}\label{ex3div}\begin{array}
         {l}\tilde{\div}_{A,10}|(1)(7)(4)(0)+\rangle_{10} 1/\sqrt{2}(|(1)(3)+
         \rangle_{10}+|(2)(1)+\rangle_{10})
         = |(1)(7)(4)(0)+\rangle_{10}\\ \times 1/\sqrt{2}
         (|(1)(3)+\rangle_{10} |(10)(03)+(11)\rangle_{13}+
         |(2)(1)\rangle_{10}|(03)(19)+(14)\rangle_{21}),\end{array}
         \end{equation} shows entanglement of base $10$ and $13$
         states with base $10$ and $21$ states. The base
         entanglement can be removed by converting both
         $|(10)(03)+(11)\rangle_{13}$ and
         $|(03)(19)+(14)\rangle_{21}$ to base $13\times 21 = 273$
         base states. The state entanglement with the divisor component
         states still remains as does the base change.

         There are two ways to handle this problem.
         One is to use this definition and deal with the complexity of
         base changes.  The advantages are that, with this definition,
         the axioms for $Ra$ for division are simple in that $Ra$ is
         closed under division.

         The other option is to use a definition of division to
         arbitrary accuracy. This definition, which is much used
         in actual computations by computers, relies on the fact
         that one can always stop the generation of an infinite
         string of digits at any point. This is easiest to see in
         computing inverses. For example $1/3$ in the decimal base
         is an infinite string of $3s$ to the right of the decimal
         point. Limiting the string to $\ell$ $3s$ to the right of the
         decimal point is division to accuracy $\ell$ as the result
         is accurate to $10^{-\ell}.$

         This definition will be used here because there is no
         change of base. This greatly simplifies the
         treatment.  The difficulty is the complexity of the
         axioms that express the concept of "division to arbitrary
         accuracy".  However since there is no emphasis on axiomatic
         details here, this is not much of a problem.

         The definition of an operator $\tilde{\div}_{A,k,m,\ell}$
         for each $\ell$ is similar to that for $\tilde{O}_{A,k,m}$
         in Eq. \ref{arithops}:\begin{equation}\label{divAl}\begin{array}{l}
         \tilde{\div}_{A,k,m,\ell}|\g,(m,h),s\rangle_{k}|\gp,(m,h),\sp
         \rangle_{k} \\ \hspace{1cm}=|\g,(m,h),s\rangle_{k}|\gp,(m,h'),
         \sp\rangle_{k} |\gpp,(m,h''),\spp\rangle_{k,\div_{\ell}}.\end{array}
         \end{equation} The quotient state\begin{equation}\label{qdivA}
         |\gpp,(m,h''),\spp,l'',u''\rangle_{k,\div_{\ell}}=|(m,h'')
         (\g,s\div_{A,\ell}\gp, \sp\rangle_{k}\end{equation} satisfies
         the condition, $$\mbox{If $l^{\p\p}<m-\ell,$ then
         $\spp(j,h'')=0$ for all $j<m-\ell.$}$$ In other words,
         the quotient state is accurate to \mbox{$|+,(m,h''),
         0_{[m-\ell+1,m]}1_{m-\ell}\rangle_{k}.$} Since the $\tilde{N}$
         eigenvalue of this state is $k^{-\ell},$ this is equivalent
         to saying that the result is accurate to $k^{-\ell}.$ More details on
         $\tilde{\div}_{A,k,m,\ell}$ are given in \cite{BenRRCNQT,BenRCRNQM}.

         \section{Space of Quantum Theory Representations of
         Numbers}\label{SQTRN}
         So far a quantum representation of numbers
         as states of finite strings of base $k$ qukits $q_{k}$ has
         been described. The states are elements of a basis set
         $\mathcal{B}_{k,(m,h)}$
         that spans a Fock space $\mathcal{F}^{X}_{k,(m,h)}$ where
         $X=N,I,$ or $Ra.$ Arithmetic relations and operations
         are, in general, defined on $n-tuples$ of $q_{k}$ string
         states.  These states are elements of $n-fold$ tensor
         products of $\mathcal{F}^{X}_{k,(m,h)}.$ If $S=\{(m,h)\}$ is a finite
         subset of  $n$ pairs of integers where the values of $h$ are all
         different then an $n-tuple$ operation or relation would be
         defined on $\mathcal{F}^{X}_{k,S}$ where\begin{equation}\label{fockS}
         \mathcal{F}^{X}_{k,S}=\bigotimes_{(m,h)\epsilon
         S}\mathcal{F}^{X}_{k,(m,h)}.\end{equation}

         If the definitions of the relations and operations are insensitive
         to the values of $(m,h),$ then the appropriate domain of
         definition would be the space $\mathcal{F}^{X}_{k,n}
         =\bigoplus_{S:|S|=n}\mathcal{F}^{X}_{k,S}$ or the space
         $\mathcal{F}^{X}_{k}=\bigoplus_{S}\mathcal{F}^{X}_{k,S}.$
         Here the sum is over all finite subsets $S$ of $I\times I.$

         From this one sees that $\mathcal{F}^{X}_{k,(m,h)}$ is the
         basic space as it contains states of single strings of
         $q_{k}.$  These states and the
         space, $\mathcal{F}^{X}_{k,(m,h)}$, are parameterized by
         three parameters, $k,m,h$ that represent the number base,
         and the location of a string in $I\times I.$ $m$ is the
         location of the sign and $"k-al"$ point in a string and
         $h$ is the location of a string.\footnote{ Note that the
         two dimensions in $I\times I$
         are treated differently. This is a consequence of the fact
         that one cannot have product states of two $q_{k}$ strings
         with the same value of $h$ but different values of $m$.
         One does not want overlapping strings
         of $q_{k}'s$ with the same $k$ because one does not know which
         $q_{k}$ belong to which string in the overlap regions. This
         problem does not exist for overlapping $q_{k}$ and $q_{k'}$
         strings where $k'\neq k$.  Such overlaps are allowed.}

         There remains a degree of freedom that should be accounted
         for.  This is the freedom of gauge fixing or basis choice
         for the states of each $q_{k}$. This is represented here
         by a gauge fixing function, $g,$ that chooses a basis set
         for the $k$ dimensional Hilbert space of states for each
         integer pair in $I\times I.$ That is, $g(k,j,h)$ is the
         basis set  of the states of $q_{k}$ that span the $k$
         dimensional Hilbert state space $\mathcal{H}^{k}_{j,h}$
         at site $j,h.$

         The choice of a gauge or basis set for each qukit at each
         location is often referred to as a choice of quantization
         axis at each site $(j,h).$ Physically this is represented by
         a vector field on $I\times I$, such as a magnetic field.
         It is also described as a moving frame \cite{Mack}.
         As defined $g$ is a function from $N\geq 2\times I\times I\times$
         to basis sets of finite dimensional Hilbert spaces where
         $g(k,j,h)$ is a basis for $\mathcal{H}^{k}_{j,h}.$

         The inclusion of $g$ as a component of the parameter space
         is done because here it is an independent variable. No
         external fields are present to determine $g$. The function
         $g$ is also quite different from the other components of
         the space. Unlike the other components, it is not a
         property of the $q_{k}$ or $q_{k}$ strings. As a $k$
         dependent choice of a basis set for each site,\footnote{The
         $k$ dependence of $g$ can also be represented as a subscript
         as in $g_{k}(j,h)$. In this case there are many $g$ functions,
         one for each $k$.} it is essential for the
          description of arithmetic properties of the $q_{k}$
          strings.  This was seen in the previous section where the
          descriptions of arithmetic relations and operations are
          given relative to a basis set of states of qukit strings.

          These considerations indicate that a subscript $g$ should
          be added to all the basis states and arithmetic relations
          and operations discussed in Section \ref{QRNRIRN}. It was
          not done there to conform with general usage.  However it
          will be included where appropriate from now on.

        The next step is to associate a state space and basis with
        each point of the parameter space. Here this association is
        defined by the three maps\begin{equation}\label{repsp}
        (k,(m,h),g)\rightarrow(\mathcal{F}^{X}_{k,(m,h)},
        \mathcal{B}_{k,(m,h),g})\end{equation} one each for
        $X=N,I,Ra.$ The righthand site  contains two elements,
        a Fock space $\mathcal{F}^{X}_{k,(m,h)}$ and a set
        $\mathcal{B}_{k,(m,h),g}$.  This set is the product of all
        basis sets $\{g(k,j,h):j\epsilon I\}.$ It also includes a
        basis for the sign qubit at
        $(m,h).$ The elements of $\mathcal{B}_{k,(m,h),g}$ are the
        qukit string states $|\g,(m,h),s,l,u\rangle_{k,g}$ for all
        values of $\g,s,l,u.$

        To save on notation the pair $(\mathcal{F}^{X}_{k,(m,h)},
        \mathcal{B}_{k,(m,h),g})$ will be denoted by
        $\mathcal{FB}^{X}_{k,(m,h),g}.$  The set of all
        $\mathcal{FB}^{X}_{k,(m,h),g}$ for all values of the
        parameters in the parameter set is the space of
        representations referred to in the title of this paper.

        \subsection{Transformations}
        So far one has a representation space with elements
        $\mathcal{FB}^{X}_{k,(m,h),g}$ of the space parameterized
        by the $4-tuples$ $((m,h),k,g).$ It is of interest to
        investigate transformations on the parameter set and their
        correspondents on the representation space.

\begin{figure}[h]\begin{center}
           \resizebox{150pt}{150pt}{\includegraphics[300pt,150pt]
           [540pt,390pt]{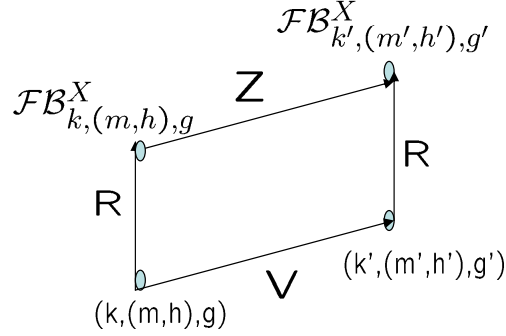}}\end{center}
           \caption{Schematic Diagram of the Commutation Condition for
           Parameter Set and Induced Representation Space
           Transformations. If $V((k,(m,h),g))=(k',(m',h'),g'),$
           and $R((k,(m,h),g))=\mathcal{FB}^{X}_{k,(m,h),g},$ and
           $Z(\mathcal{FB}^{X}_{k,(m,h),g})=\mathcal{FB}^{X}_{k',
           (m',h'),g'},$ then $V,R,Z$ must satisfy
           $R(V((k,(m,h),g)))=Z(R((k,(m,h),g))).$}
           \label{NP2} \end{figure}

        To this end one notes that transformations
        $(k,(m,h),g)\rightarrow (k',(m',h'),g')$ in
        the parameter space induce transformations $\mathcal{FB}^{X}
        _{k,(m,h),g}\rightarrow \mathcal{FB}^{X}_{k',(m',h'),g'}$
        in the representation space. This is shown schematically in
        Fig. \ref{NP2}.  The transformations are of three types,
        translations, base changes, and gauge transformations.

         \subsection{Translations and Base Changes}\label{TBC}
        There are two translation operators, $\tilde{T}_{1}$ and
        $\tilde{T}_{2}.$  These operators shift the $q_{k}$ string
        one step by shifting either the first or the second integer
        parameter by $+1.$ One has
         \begin{equation}\label{defT}\begin{array}{l}
         \tilde{T}_{1}|\g,(m,h),s,(l,h),(u,h)\rangle_{k,g} =
         |\g,(m+1,h),\sp,(l+1,h),(u+1,h)\rangle_{k,g} \\
         \tilde{T}_{2}|\g,(m,h),s,(l,h),(u,h)\rangle_{k,g} =
         |\g,(m,h+1),s,(l,h+1),(u,h+1)\rangle_{k,g} \end{array}
         \end{equation} where $\sp(j+1,h)=s(j,h)$ for
         $l\geq j\geq u$ and $s,s'$ are independent of $h$.
         $\tilde{T}_{1}$ and $\tilde{T}_{2}$ can be regarded as
         step translation operators in the direction along the
         $q_{k}$ string or at right angles to the string.

         $\tilde{T}_{1}$ and $\tilde{T}_{2}$ are unitary.
         They also have the property that for
         each $k$ and $(m,h)$\begin{equation}\label{TFX}\begin{array}{c}
         \tilde{T}_{1}\mathcal{FB}^{X}_{k,(m,h),g} =
         \mathcal{FB}^{X}_{k,(m+1,h),g} \\ \tilde{T}_{2}\mathcal{FB}^
         {X}_{k,(m,h),g} =\mathcal{FB}^{X}_{k,(m,h+1),g}.\end{array}\end{equation}
          Also the states $\tilde{T}_{1}|\g,m,s,l,u\rangle_{k,g}$ and
          $\tilde{T}_{2}|\g,m,s,l,u\rangle_{k,g}$ represent the same
          number in $\mathcal{FB}^{X}_{k,(m+1,h),g}$ and in
          $\mathcal{FB}^{X}_{k,(m,h+1),g}$ as $|\g,m,s,l,u\rangle_{k,g}$
          does in $\mathcal{F}^{XB}_{k,(m,h),g}.$

         Transformations that change bases are more complex,
         especially for rational number states. For each pair
         $k,\kp$ of bases, let $\tilde{W}_{\kp,k}$ denote the
         operator that changes any  state from base $k$ to one in base
         $\kp.$ The operator must satisfy the requirement that for
         each  state $|\g,(m,h),s\rangle_{k,g}$ in $\mathcal{FB}^{X}_{k,
         (m,h),g}$ on which $\tilde{W}_{\kp,k}$ is defined,
         $\tilde{W}_{\kp,k}|\g,(m,h),s\rangle_{k,g},$ as a state in
         $\mathcal{FB}^{X}_{\kp,(m,h),g},$ represents the same number as
         $|\g,(m,h),s\rangle_{k,g}$ does in $\mathcal{FB}^{X}_{k,(m,h),g}.$
         Note that if $\tilde{W}_{\kp,k}$ is defined for one value of
         $(m,h)$ it is defined for all values of $(m,h)$. For notational
         simplicity, $l$ and $u$ have been suppressed.

         It is worth giving two examples of base changes.  The symbol
         representation used for the division examples, Eqs. \ref{ex2div},
         \ref{ex3div}, is used here. The base $37$ integer
         $(23)(35)(0)._{37},$ also represented as $23\times
         37^{2}+35\times 37+0$,  has a base $10$ representation as
         $(3)(2)(7)(8)(2)._{10}.$ As another example, the rational number
         $(23)(35)(0).(1)_{37},$ where the subscript denotes
         the base $37,$ has an equivalent representation as $23\times
         37^{2}+35\times 37 +0+1/37.$  This number does not have a
         finite string representation in base $k=10.$

         As a transformation operator on the representation space
         $\tilde{W}_{\kp,k}$ is very different from
         $\tilde{T}_{1}$ and $\tilde{T}_{2}$ in that these
         operators, and the to-be-described gauge changing operator,
         are defined on states of individual $q_{k}.$
         $\tilde{W}_{\kp,k}$ is defined on states in
         $\mathcal{FB}^{X}_{k,(m,h)}$ as it is a string state
         changing operator. It is not defined on states of
         individual  $q_{k}$. This will become quite evident in the
         following.

         The requirement that $\tilde{W}_{\kp,k}|\g,m,s,l,u\rangle_{k,g}
         =|\g,m,\sp,l^{\p},u^{\p}\rangle_{\kp,g}$ represent the same
         number as does $|\g,m,s,l,u\rangle_{k}$ is not trivial. It
         means that both states must have exactly the same numerical
         properties, expressed as  theorems derivable from the
         set of axioms for the number type being considered. A rigorous
         proof of this would require proving that the axiomatic
         properties of the basic arithmetic relations and operations
         are preserved, and that the logical deduction rules for
         obtaining theorems from axioms are invariant. A necessary
         condition for this  to be true is the requirement that the
         properties of the basic arithmetic relations and
         operations, as given by the axioms for the number type being
         considered, are preserved.

         For example, for $X=I$ the basic arithmetic relations and
         operations are $=_{A,k,m,g},\leq_{A,k,m,g}, +_{A,k,m,g},\times_{A,k,m,g}$.
         The requirement that $\tilde{W}_{\kp,k}$ preserves these
         properties for the relations is expressed by ($h$ is suppressed here)
         \begin{equation}\label{arithprop}\begin{array}{l}|\g,m,s,l,u\rangle_{k,g}
         =_{A,k,m,g}|\gp,m,\sp,l,u^{\p}\rangle_{k,g}\\\rightarrow \tilde{W}_{\kp,k}
         |\g,m,s,l,u\rangle_{k,g}=_{A,\kp,m,g}\tilde{W}_{\kp,k}|\gp,m,\sp,l,u^{\p}
         \rangle_{k,g},\\|\g,m,s,l,u\rangle_{k,g}\leq_{A,k,m,g}|\gp,m,\sp,l,
         u^{\p}\rangle_{k,g} \\ \rightarrow \tilde{W}_{\kp,k}|\g,m,s,l,u
         \rangle_{k,g}\leq_{A,\kp,m,g}\tilde{W}_{\kp,k} |\gp,m,\sp,l,u^{\p}
         \rangle_{k,g}\end{array}\end{equation}. The requirement for the
         arithmetic operations $+_{A,k,m,g},\times_{A,k,m,g}$ and
         $-_{A,k,m,g}$ can be expressed by
         \begin{equation}\label{arops}\begin{array}{l}\tilde{W}_{k',k}
         |(\g,m,s,l,u)O_{A,k,m,g}(\g',m,s',l',u')\rangle_{k,g}\\\mbox{}
         \hspace{1cm}=_{A,k'm,g}|(W_{k',k}(\g,m,s,l,u))
         O_{A,k',m,g}(W_{k',k}(\g',m,s',l',u)')\rangle_{k',g}\end{array}
         \end{equation} where $O$ stands for $+,$ $\times$ and $-.$
         This equation says that transforming the state resulting
         from carrying out the operation $O_{A,k,m,g}$ gives a state
         that is arithmetically equal to the state resulting from
         carrying out $O_{A,k',m,g}$ on the transformed states. The
         states appearing in the above must satisfy the
         restrictions on $s,s'$ for integer states: If $l\leq j< m,$
         then $s(j)=0$ and $\sp(j)=0.$

         The operator $\tilde{W}_{\kp,k}$ has different properties
         for $X=Ra$ than for $X=N,I.$ For $X=N,I,$
         $\tilde{W}_{\kp,k}$ is defined on all of $\mathcal{FB}^{X}
         _{k,(m,h),g}$ and is unitary. In this case, for all $\kp$,
         \begin{equation}\label{Wkpk} \tilde{W}_{\kp,k}\mathcal{FB}^{X}_{k,(m,h),g}
         =\mathcal{FB}^{X}_{\kp,(m,h),g}.\end{equation} Also $\tilde{W}_{k,k}$
         is the identity on $\mathcal{FB}^{X}_{k,(m,h),g},$ and
         $\tilde{W}^{\dag}_{\kp,k}=\tilde{W}_{k,\kp}.$ Also the operators
         $\tilde{W}_{\kp,k}$ have the group multiplication property in
         that\begin{equation}\label{Wkkpkpp}\tilde{W}_{\kpp,k}=
         \tilde{W}_{\kpp,\kp}\tilde{W}_{\kp,k}\end{equation} for all
         triples $\kpp,\kp,k.$

         Note here that $\tilde{W}_{\kp,k}$ is defined on the pair
         $\mathcal{FB}^{X}_{k,(m,h),g}$ and not just on
         $\mathcal{F}^{X}_{k,(m,h)}.$  This expresses the
         requirement that $\tilde{W}_{\kp,k},$ as an operator on the
         Fock space $\mathcal{F}^{X}_{k,(m,h)},$ takes basis states
         in $\mathcal{B}^{X}_{k,(m,h),g}$ to basis states in
         $\mathcal{B}^{X}_{k',(m,h),g}.$

         For $X=N,I,$ $\tilde{W}_{\kp,k}$ can be
         replaced by a simpler index independent operator
         $\tilde{W}$ on $\mathcal{FB}^{X}_{k,g}=
         (\mathcal{F}^{X}_{k},\mathcal{B}_{k,g})$
         which increases the base by $1$ unit. For each base $k,$
         \begin{equation}\label{W} |\g,m,\sp,l',\up\rangle_{k+1,g}=
         \tilde{W}|\g,m,s,l,u\rangle_{k,g} \end{equation}is a
         state of a string of base $k+1$ qukits. It represents the
         same number as does $|\g,m,s,l,u\rangle_{k},$ which is a
         state of a string of base $k$ qukits. Note that $\tilde{W}$
         is not unitary.   It is an isometry or unilateral shift
         \cite{Halmos} because of the lower limit
         of $k\geq 2.$ One sees that iteration of the action of
         $\tilde{W}$ on $\mathcal{FB}^{X}_{2,g}$ generates all spaces
         $\mathcal{FB}^{X}_{k,g}.$ However, $\tilde{W}^{\dag}\mathcal
         {FB}^{X}_{2,g}=0.$ It follows that $\tilde{W}$ is the
         generator of a semigroup of transformations on
         $\mathcal{FB}^{X}_{g}.$

         This simple description does not extend to
         $\mathcal{FB}^{Ra}_{k,g}$ as $\tilde{W}$ is not defined on
         $\mathcal{FB}^{Ra-I}_{k,g},$  the noninteger part of
         $\mathcal{FB}^{Ra}_{k,g}.$ Instead one has to restrict
         consideration to the indexed base change operators
         $\tilde{W}_{\kp,k}.$

         The domain and range of $\tilde{W}_{\kp,k}$ depend on the
         relation between the prime factors
         of $k$ and $\kp.$ If $k$ and $k^{\p}$ have no common prime
         factors, then $\tilde{W}_{\kp,k}$ is not defined on the
         noninteger part $\mathcal{FB}^{Ra-I}_{k,g}$ of
         $\mathcal{FB}^{Ra}_{k,g}$. It is defined on the integer subspaces
         of ${\mathcal FB}^{Ra}_{k,g}$ and ${\mathcal FB}^{Ra}_{\kp,g}$
         and is an arithmetic isomorphism (a unitary operator that
         preserves arithmetic relations and operations) on the subspaces.
         It satisfies \begin{equation}\label{qmWkk1} \tilde{W}_{\kp,k}
         {\mathcal FB}^{Ra}_{k,g}|_{int}=\mathcal{FB}^{Ra}_{\kp,g}|_{int}.\end{equation}

           For cases in which $k$ and $k_{\p}$ have prime factors
           in common, the domain and range of $\tilde{W}_{\kp,k}$
           includes some noninteger states in $\mathcal{FB}^{Ra}_{k,g}.$
           The different cases can be summarized as
           follows:  Let $PF(k)$ denote the prime factors of $k$.
           Then\begin{equation}\label{domranW}
           \begin{array}{l}\mbox{If $PF(k)\subset PF(k')$ then
           $\tilde{W}_{\kp,k}\mathcal{FB}^{Ra}_{k,g}\subset
           \mathcal{FB}^{Ra}_{k',g},$}\\ \mbox{If $PF(k)\supset PF(k')$ then
           $\tilde{W}_{\kp,k}\subset\mathcal{FB}^{Ra}_{k,g}=
           \mathcal{FB}^{Ra}_{k',g},$}\\ \mbox{If $PF(k),PF(k')$ each have elements
           not in the other and}\\ \mbox{}\hspace{0.8cm}\mbox{share
           elements in common, then } \tilde{W}_{\kp,k}\subset\mathcal{FB}^{Ra}_{k,g}=
           \subset\mathcal{FB}^{Ra}_{k',g},\\ \mbox{If $PF(k)= PF(k')$ then
           $\tilde{W}_{\kp,k}\mathcal{FB}^{Ra}_{k,g}=\mathcal{FB}^{Ra}_{k',g}.$}
           \end{array}\end{equation}In  the above $\subset\mathcal{FB}^{Ra}_{k,g}$
           denotes a subspace of $\mathcal{FB}^{Ra}_{k,g}.$ In all these cases,
           if the state $|\g,(m,h),s,l,u\rangle_{k,g}$ is in the domain of
           $\tilde{W}_{\kp,k},$ then the base $\kp$ state,
           $\tilde{W}_{\kp,k}|\g,(m,h),s,l,u\rangle_{k,g},$
           represents the same rational number as does
           $|\g,(m,h),s,l,u\rangle_{k,g}.$

           The case where $PF(k)=PF(k')$ is of special interest
           because for each $k$ there is a smallest $k'$ that has
           the same prime factors as $k$.  If\begin{equation}\label{kprfa}
           k=p_{j_{1}}^{h_{1}}\cdots p_{j_{n}}^{h_{n}},\end{equation}
           then the smallest $\kp$ is given by \begin{equation}\label{kpprfa}
           k^{\p}=p_{j_{1}}\cdots p_{j_{n}}. \end{equation} Here
           $p_{j_{a}}$ for $a=1,2,\cdots,n$ is the $j_{a}th$ prime number.

            A special example of this consists of the sets of $k$ that
            have the first $n$ primes as their factors for
            $n=1,2,\cdots.$  Define $k_{n}$ by
            \begin{equation}\label{kbase} k_{n}=p_{1}p_{2}\cdots
            p_{n}=2\times 3\times\cdots\times p_{n}.\end{equation}
            Then the basis states $|\g,m,s,l,u\rangle{k_{n},g}$ in
            ${\mathcal FB}_{k_{n},g}$ represent the
            same numbers as do the basis states in
            $\mathcal{FB}_{\kp,g}$ where $\kp$ is any base that
            has the same prime factors as $k_{n}$.

            As might be expected the group multiplication properties
            of $\tilde{W}_{k',k}$ depend on the relation between the
            prime factors of $k$ and $k'.$ Let $[k]$ be the set of
            all $k'$ that have the same prime factors as $k$. Then,
            for all $k,k',k''$ in $[k]$, \begin{equation}\label{Wmult1}
           \tilde{W}_{\kpp,k}=\tilde{W}_{\kpp,\kp}\tilde{W}_{\kp,k}.
           \end{equation} and\begin{equation}\label{Wdag}
           \tilde{W}^{\dag}_{k',k}=\tilde{W}_{k,k'}.\end{equation}
           Note also that $\tilde{W}_{k,k}$ is the identity
           on $\mathcal{F}^{Ra}_{k}.$

           These properties of the $\tilde{W}_{k',k}$ for different $k,k'$ mimic
           the corresponding properties of subsets of rational numbers expressed
           as finite digit strings in any base $k.$ To see this one notes that
           the set, $Ra_{k},$ of rational numbers expressible by states of
           finite strings of base $k$ qukits is also representable by the
           set of numbers $i/k^{n}$ for $n=0,1,\cdots$ where $i$ is
           any integer such that if $n>0,$ then $i$ does not have
           $k$ as a factor. This follows from the observation that, for
           any rational number, one can always shift the $k-al$ point
           to the right hand end by multiplying by a power of $k$.
           For example, $97.31=9731.\times 0.01$.

           One sees that the relations given for the $\tilde{W}_{\kp,k}$
           also apply to the different $Ra_{k}.$ If $k$ and $\kp$ have no
           common prime factors, then the integers are the only rational
           numbers that $Ra_{k}$ and $Ra_{k^{\p}}$ have in common.
           If $k$ has prime factors not in $k^{\p}$ and all prime
           factors of $k^{\p}$ are factors of $k,$ then $Ra_{k^{\p}}
           \subset Ra_{k}.$ If $k$ and $k^{\p}$ have the same
           prime factors then $Ra_{k}=Ra_{k^{\p}}.$

          \subsection{Gauge Transformations}\label{GT}

          So far all components of the  transformation
          $\mathcal{FB}^{X}_{k,(m,h),g}\rightarrow\mathcal{FB}^{X}_{k',(m',h'),g'}$
          have been treated except the change of gauge or basis  from $g$ to
          $g'.$ This is done by means of  gauge transformations
          $U_{k}.$ Here $U_{k}$ is defined as a $U_{1}\times SU(k)$
          valued function\begin{equation}\label{Gaugedef}
          U_{k}:I\times I\rightarrow U_{1}\times SU(k)\end{equation}on
          $I\times I.$ $U_{k}$ is \begin{equation}
          \label{GlLo}\begin{array}{c}\mbox{global if $U_{k}(i,j)$ is
          independent of $(i,j)$}\\\mbox{local if $U_{k}(i,j)$
          depends on $(i,j).$}\end{array}\end{equation}

          Gauge transformations are different from the base change and
          shift transformations in that they have no counterpart in
          classical representations of rational numbers as finite strings of
          digits in a base $k$.  Unlike the other transformations,
          which apply to both classical kit strings and quantum
          qukit strings, gauge freedom and the associated
          transformations from one gauge to another are strictly
          quantum theoretical. This follows from the observation
          that the choice of a gauge corresponds to the choice of a basis
          in the $k$ dimensional Hilbert space of states for each qukit
          integer location.

          In many situations the choice of gauge is fixed. It plays
          no role in the representation of states and dynamics of
          qukits. However there are other situations, such as those
          occurring in quantum cryptography  \cite{Nielsen,Crypt,Enk}
          where rotations of the axis, or gauge transformations,
          play an important role. Gauge transformations also play a
          role in the construction of decoherence free subspaces for
          reference frame changes in quantum information
          \cite{Byrd,Kempe,Bartlett} and in gauge theories \cite{Gross,Montavy}.

          The effect of gauge transformations $U_{k}$ on a state
          $|\g,m,s,l,u\rangle_{k,g}$ of a string of base $k$ qukits
          is given by \begin{equation}\label{Urast}\begin{array}{l}
          U_{k}|\g,m,s,l,u\rangle_{k,g} =\cd_{\g,m} U_{k}(u)(\ad_{k})_{s(u),u}
          \cdots U_{k}(l)(\ad_{k})_{s(l),l}|0\rangle \\ \hspace{1cm}=
          (\cd_{\g,m}((\ad_{k})_{U_{k}(u)})_{s(u),u}\cdots((\ad_{k})_
          {U_{k}(l)})_{s(l),l}|0\rangle\end{array}\end{equation}
          where\begin{equation}\label{adU}\begin{array}{c}
          ((\ad_{k})_{U_{k}(j)})_{\alpha,j}=U_{k}(j)(\ad_{k})_{\alpha,j}=
          \sum_{\beta}U_{k}(j)_{\alpha,\beta}(\ad_{k})_{\beta,j}\\
          ((a_{k})_{U_{k}(j)})_{\beta,j}=(a_{k})_{\beta,j}U^{\dag}_{k}(j)=
          \sum_{\alpha}U^{*}_{k}(j)_{\alpha,h}a_{\alpha,j}\end{array}\end{equation}
          These results are based on the representation of $U_{k}(j)$
          as \begin{equation}\label{Uexpand} U_{k}(j)=\sum_{\alpha,\beta}
          (U_{k}(j))_{\alpha,\beta}(\ad_{k})_{\alpha,j}(a_{k})_{\beta,j}.
          \end{equation} The parameter $h$ is suppressed in the above as
          it is the same for each qukit A-C operator. Thus $U_{k}(j)$
          denotes $U_{k}(j,h)$ which is an element of $U(k)=U(1)\times SU(k)$
          and $((\ad_{k})_{U_{k}(j)})_{\alpha,j}$ denotes $((\ad_{k})_{U_{k}
          (j,h)})_{\alpha,(j,h)}.$ Also from now on the $m$ subscript will
          be often suppressed unless it it needed to help in understanding.
          Note that, in the interest of simplicity, gauge transformations for
          the sign qubit are not considered here.  Adding them by including
          $U_{2}(m)$ as an element of $U_{1}\times SU(2)$ adds nothing new.

          The need for the subscript $g$ is evident now in that one
          has for any state $|\g,m,s,l,u\rangle_{k,g}$
          \begin{equation}\label{Ugg'}U_{k}|\g,m,s,l,u\rangle_{k,g}
          = |\g,m,s,l,u\rangle_{k,g'}.\end{equation} Without the subscript
          change $g\rightarrow g'$ there would be no way to show
          that the two states are different.  Note that the state
          $|\g,m,s,l,u\rangle_{k,g'}$ is a linear superposition of
          the states $|\g,m,s,l,u\rangle_{k,g}$\begin{equation}\label{supUk}
          |\g,m,s,l,u\rangle_{k,g'}=\sum_{\g',s'}|\g',m,s',l',u'\rangle_{k,g}
          \langle\g',m,s',l',u'|U_{k}|\g,m,s,l,u\rangle_{k,g}.\end{equation}
          Here the $s'$ sum includes a sum over $l',u'.$

          A single qubit example of these equations is given by
          $|i\rangle_{g'}=U_{2}|i\rangle_{g}$ for $i=0,1$  where
          $U_{2}=1/\sqrt{2}\left| \begin{array}{ll}1 & 1 \\-1 & 1
          \end{array}\right|.$  Then $|1\rangle_{g'}=|+\rangle_{g}=1/\sqrt{2}
          (|1\rangle_{g}+|0\rangle_{g})$ and $|0\rangle_{g'}=|-\rangle_{g}
          =1/\sqrt{2}(|1\rangle_{g}-|0\rangle_{g}).$

          Arithmetic relations and operators transform in the
          expected way. For the relations one defines
         $=_{A,k,g'}$ and $\leq_{A,k,g'}$ by\begin{equation}\label{=AG}
         \begin{array}{c}=_{A,k,g'}:= (U_{k}=_{A,k,g}U_{k}^{\dag}) \\
         \leq_{A,k,g'}:= U_{k}\leq_{A,k,g} U_{k}^{\dag}.
         \end{array}\end{equation} These relations express the fact
         that $U_{k}|\g, s\rangle_{k,g} =_{A,k,g'}U_{k}|\gp,\sp
         \rangle_{k,g}$ if and only if $|\g,s\rangle_{k,g} =_{A,k,g}
         |\gp,\sp\rangle_{k,g}.$ Also $U_{k}|\g,s\rangle_{k,g}
         \leq_{A,k,g'}U_{k}|\gp,\sp\rangle_{k,g}$ if and only if
         $|\g,s\rangle_{k,g}\leq_{A,k,g}|\gp,\sp\rangle_{k,g}.$ Here
         and below the subscripts $g$ and $g'$ have been added to
         denote which basis is used to define the arithmetic relations,
         operations, and number basis sets.

         For any of the operations $\tilde{O}_{A,k,g}$ where $O=+,-,\times,
         \div_{\ell},$ one defines $\tilde{O}_{A,k,g'}$
         by \begin{equation}\label{AopsAU}\tilde{O}_{A,k,g'}:=(U_{k}\times
         U_{k}\times U_{k})\tilde{O}_{A,k,g}(U_{k}^{\dag}\times U_{k}^{\dag}).
         \end{equation}Then
         \begin{equation}\label{opsAUA}\begin{array}{l}\tilde{O}_{A,k,g'}
         |U_{k}(\g,s)\rangle_{k,g'}|U_{k}(\gp,\sp)\rangle_{k,g'})\\
         \mbox{}\hspace{1cm}=|U_{k}(\g,s)\rangle_{k,g'}|U_{k}(\g',s')
        \rangle_{k,g'}|(U_{k}(\g,s))O_{A,k,g'}(U_{k}(\g's'))\rangle_{k,g'}
        \end{array}\end{equation} where\begin{equation}\label{Ugs}
        \begin{array}{l}|U_{k}(\g,s)\rangle_{k,g'}=U_{k}|\g,s\rangle_{k,g}
        \\ |U_{k}(\g',s')\rangle_{k,g'}=U_{k}|\g',s'\rangle_{k,g}\\|(U_{k}(\g,s))
         O_{A,k,g'}(U_{k}(\g's'))\rangle_{k,g'}=U_{k}|(\g,s)O_{A,k,g}
         (\g'.s')\rangle_{k,g}.\end{array}\end{equation} These equations,
         which are based on Eq. \ref{arithops}, show the transformations
         of the basic arithmetic operations on $g$ gauge states to those
         on $g'$ gauge states.

         The different number of $U_{k}$ and $U^{\dag}_{k}$ factors
         arises because the operations $\mathcal{O}_{A,k,g}$, acting
         on two $q_{k}$ string states, create a third $q_{k}$ string
         state. For this paper it is immaterial whether this refers
         to creation of new $q_{k}'s$ or to transfer from some supply of
         $q_{k}'s$. In the latter case the total number of $q_{k}$ is
         preserved.

         \subsection{Commutation Relations}\label{CR}
         It is of interest to investigate the commutation relations
         between $\tilde{T}_{1},\;\tilde{T}_{2},\;\tilde{W}_{\kp,k},$ and $U_{k}.$ The
         simplest is between $\tilde{T}$ and $\tilde{W}_{k,\kp}$ in
         that these commute: \begin{equation}\label{TWcomm}
         \tilde{T}_{i}\tilde{W}_{\kp,k}-\tilde{W}_{\kp,k}\tilde{T}_{i}=0
         \end{equation} for $i=1,2.$ Because $\tilde{W}_{\kp,k}$ is not defined for
         all $k,\kp,$ this makes sense only in cases where this
         operator is defined. Also $\tilde{T}_{1}$ commutes with $\tilde{T}_{2}$

         The commutation relation for $\tilde{T}_{i}$ and $U_{k}$ is
         straightforward. One has \begin{equation}\label{TUcomm}
         \tilde{T}_{i}U_{k}-U^{i}_{k}\tilde{T}_{i}=0.\end{equation} Here
         $U^{i}_{k}$ for $i=1,2$ is defined by $U^{1}_{k}(j+1,h)=U_{k}(j,h)$
         and $U^{2}_{k}(j,h+1)=U_{k}(j,h).$ This shows that $\tilde{T}_{i}$
         and $U_{k}$ commute if and only if $U_{k}$ is global in the $ith$ index.

         The problems come when one attempts to give a commutation
         relation between $\tilde{W}_{\kp,k}$ and $U_{k}$. It
         appears that this is possible if and only if $\kp$ is a power of
         $k,$ such as $\kp=k^{n}.$ The reason this works is that
         there is a one-one map from the set of all $n-tuples$ of
         base $k$ digits onto the set of base $\kp$ digits.  As a
         simple example let $k=2$ and $\kp =8.$  Then there is a
         one-one map between the triples $000,001,\cdots,111$ and
         $0,1,\cdots,7.$ For $\kp =k^{n}$ the commutation relation is
         \begin{equation}\label{WUcomm}\tilde{W}_{\kp,k}U_{k}-U^{\p}
         _{\kp}\tilde{W}_{\kp,k}=0.\end{equation} Here each element
         of $U^{\p}_{\kp}$ is a product of successive $n-tuples$ of
         elements of $U_{k}.$ One has ($h$ is suppressed as it is the
         same everywhere)\begin{equation}\label{UUp}
         \begin{array}{l}U^{\p}_{\kp}(m-j)=\begin{array}{l}U_{k}(m-n(j-1)-1)\times
         U_{k}(m-n(j-1)-2)\times \\ \hspace{0.5cm} \cdots\times U_{k}(m-nj)
         \mbox{ for $j\geq 1$}\end{array} \\ U^{\p}_{\kp}(m+j)=\begin{array}{l}
         U_{k}(m+n(j+1)-1)\times U_{k}(m-n(j+1)-2)\times\\ \hspace{0.5cm}
         \cdots\times U_{k}(m+jn)\mbox{ for $j\geq 0$}.\end{array}\end{array}
         \end{equation}

         Even though there do not seem to be commutation relations between
         $U_{k}$ and $\tilde{W}_{k',k}$ for arbitrary $k$ and $k',$ one
         can always use the gauge
         transformations to define transformed base change operators.
         For any pair, $U_{k},\; U_{\kp},$ of gauge transformations,
         the map $(\tilde{W}_{U',U})_{\kp,k},$ defined by
         \begin{equation}\label{UkpWUk}(\tilde{W}_{U',U})_{\kp,k}
         =U_{\kp}\tilde{W}_{\kp,k}U^{\dag}_{k},\end{equation} is a number
         preserving map between gauge transformed states just as
         $\tilde{W}_{\kp,k}$ is between the original states. One has
         \begin{equation}\label{WUkpk}\begin{array}{l}(\tilde{W}_{U',U})_{\kp,k}
         U_{k}|\g,m,s,l,u\rangle_{k,g}\\ \hspace{0.5cm}= U_{\kp}|\g,m,\sp,
         \lp,\up\rangle_{\kp,g}=U_{k'}\tilde{W}_{k',k}|\g,m,s,l,u\rangle_{k,g}.
         \end{array}\end{equation} Here $U_{\kp}|\g,m,\sp,\lp,\up\rangle_{\kp,g}$
         represents the same number in the gauge transformed $\kp$ basis as
         $U_{k}|\g,m,s,l,u\rangle_{k,g}$ does in the gauge transformed $k$
         basis. Note that the restrictions on the
         domain and range for $\tilde{W}_{k',k}$ in the original
         representation apply to the domain and range of
         $(\tilde{W}_{U',U})_{\kp,k}$ in the transformed
         representation.

         The meaning of the statements "...the same number as..." is
         based on the transformed basic arithmetic relations and
         operations, given by Eqs. \ref{=AG}-\ref{opsAUA}, in both
         the $k$ and $\kp$ bases. The number represented by the
         state $U_{k}|\g,m,s,l,u\rangle_{k,g}$ is determined by its
         properties relative to the transformed relations and
         operations $\tilde{=}_{z},\tilde{\leq}_{z},\tilde{+}_{z},
         \tilde{-}_{z},$ $\tilde{\times}_{z},\tilde{\div}_{z,\ell}$
         where the subscript $z$ denotes the 4-tuple $A,k,m,g'.$
         Similarly the number represented by $U_{\kp}|\g,m,\sp,\lp,
         \up\rangle_{\kp}$ is determined by properties based on the
         basic relations where $z$ denotes the 4-tuple $A,\kp,m,g'.$
         These two numbers should be the same.

         This emphasizes that one must also transform the relations and
         operations along with the state. The reason is that as
         Eq. \ref{supUk} shows, relative
         to the untransformed relations and operations, the states
         $U_{k}|\g,m,s,l,u\rangle_{k,g}$ and $U_{\kp}|\g,m,\sp,\lp,
         \up\rangle_{\kp,g}$ are linear superpositions of states
         representing many different numbers.

          \subsection{$k=1$}
         So far all number bases have been considered except one,
         the value $k=1.$ The $k=1$ string representations are
         called unary representations.  These are not usually
         considered, because basic arithmetic operations on these numbers
         are exponentially hard. For instance the number of steps
         needed to add two unary numbers is proportional to the
         values of the numbers and not to the logarithms of the
         values. However, even though they are not used
         arithmetically, they are always present in an interesting way.

         To see this one notes that $k=1$ representations are the
         only ones that are  extensive, all others are
         representational. The representational property for
         $k\geq 2$ base states of a qukit string means that a number
         represented by a state has nothing to do with the properties
         of the string state. The number represented by the
         state, $|672\rangle,$  of a string of $3$ $q_{10}'s$ is
         unrelated to the properties of the qukits in  the state.

         Extensivity of any unary representation means that
         any collection of systems is an unary representation of a
         number that is the number of systems in the collection.
         There are many examples. A system of spins on a lattice
         is an unary representation of a number, that is the number
         of spins in the system. A gas of particles in a box is an unary
         representation of a number, that is the number of particles
         in the box. The qukit strings that play such an important role in this
         paper are unary representations of numbers, that are the
         number of qukits in the strings. A single qukit is an unary
         representation of the number $1$.

         This omnipresence of unary representations relates to
         another observation that $1$ is the only number that is a
         common factor of all prime numbers and of all numbers. So
         it is present as a factor of any base.  This ties in with
         the observation that unary representations of  rational
         numbers as a single collection of systems do not exist.\footnote{
         Pairs of unary representations will work since all
         integers can be so represented.  But rational numbers
         as integer pairs are not being considered here.}
         Natural numbers and integers are the only
         ones with unary representations. This ties in nicely with
         the observation that, for any pair $k,k',$ the domain of
         $\tilde{W}_{k',k}$ includes the integer subspace of states,
         $\mathcal{FB}^{I}_{k,m,g}$ in $\mathcal{FB}^{Ra}_{k,m,g},$
         and if $k,k'$ have no prime factors in common,
         $\mathcal{FB}^{I}_{k,m,g}$  and $\mathcal{FB}^{I}_{k',m,g}$
         are the domain and range of $\tilde{W}_{k',k}.$

         The extensivity of unary representations supports
         the inclusion of the $U(1)$ factor, Eq.
         \ref{Gaugedef}, in the definition of gauge transformations.
         A state of $(\ad_{k})_{\alpha,(i,j)}|0\rangle$ of a qukit
         in state $|\alpha\rangle$ at location $(i,j)$ is also an
         unary representation of the number $1$. Multiplication of
         this state by a phase factor $e^{i\theta_{i,j}}$ is a
         transformation that gives another state that is also an
         unary representation of the number of qukits represented by the state.

         This argument extends to states of strings of qukits. A
         phase factor associated with any state of a string of
         $q_{k}$ at sites $(l,h),\cdots (u,h)$ is a product of the
         phase factors associated with each of the $q_{k}$ in the
         string.  If $e^{i\theta_{j,h}}$ is a phase factor for
         $q_{k}$ at site $(j,h),$ then $ e^{i\Theta_{[(l,h),(u,h)]}},$
          where $\Theta_{[(l,h),(u,h)]}=\sum_{j=l}^{u}\theta_{j,h},$
          is the phase factor for any state of the string.

         As is well known, multiplying any state by a phase factor
         gives the same state as far as any physical meaning is
         concerned.  However here one can have linear superpositions
         of states of strings of $q_{k}$ both at different locations
         and of different length strings.  In these cases the phase
         factors do matter as they change the relative phase between
         the components in the superposition.

         \section{Symmetries and Invariances}\label{SI}

         So far it has been seen that the spaces of quantum
         representations of $N,I,$ and $Ra$ are parameterized by the
         location, $(m,h),$ of the qukit strings,  the base $k$,
         and the choice of gauge or basis in the $k$ dimensional
         Hilbert space of states for each  $q_{k}$ location in
         $I\times I.$ Associated with this parameterization
         are the operators on the corresponding representation
         space:  translation operators $\tilde{T}_{1},\tilde{T}_{2},$
         base change operators $\tilde{W}_{\kp,k},$ and
         (base dependent) gauge transformations $U_{k}.$

         Of interest are the symmetry or invariance aspects of
         various properties and operations for the qukit strings
         and their states. In particular, one would like to know
         which properties are invariant under all the transformation
         operators and which are not.  Invariant properties can
         also be regarded as symmetries of the space.

         One set of invariant properties are those expressed by the
         axioms and theorems for each number type. This is a direct
         consequence of the fact that each axiom and theorem
         expresses a property which is valid for all representations
         in the space.  In other words, the truth value of each
         axiom and theorem is unchanged under any of the transformations
         on the representation space. Each axiom and theorem is true
         for all representations in the space.

         Another way to express this in more physical terms is to
         call a property invariant if it is conserved as one "moves"
         the qukit string around in the representation space. This
         includes changes in the locations of the string on $I\times I$,
         changes in the base, and changes in the basis for the
         states of each qukit. However there are differences. In
         particular, the change of base corresponds to a change from
         strings of one $q_{k}$ system to strings of another $q_{k'}$ system.
         Base $k$ qukits are different from base $\kp$ qukits just as a
         spin $S$ system is different from a spin $S^{\p}$ system.

         It should be emphasized again that the invariance of axioms and
         theorems under transformations in the representation space
         is not trivial and obvious. Each representation in the space
         describes qukit strings and their states for a specific
         $I\times I$ location, $(m,h),$ a specific base, $k$, and a gauge
         choice, $g$ for the states of all qukit locations. By
         themselves, the states of these systems cannot be said to
         represent numbers of any type. The validity of the
         statement that these states represent numbers
         is based on \begin{itemize}\item defining operations and
         relations on the states of the qukit strings in terms of
         basic operations on the states of the individual qukits,
         \item proving that these relations and operations satisfy
         the axioms and theorems for the number type being considered.
         \end{itemize}

         In particular, arithmetic relations, $=_{A,k,(m,h),g},$
         $\leq_{A,k,(m,h),g},$ and operations, $\tilde{+}_{A,k,(m,h),g},
         \tilde{\times}_{A,k,(m,h),g},$ for the $N,I,$ and $Ra$ spaces,
         $\tilde{-}_{A,k,(m,h),g}$ for the $I$ and $Ra$ spaces and
         $\tilde{\div}_{A,k,(m,h),g,\ell}$ are defined in terms of
         basic properties and operations on the $q_{k}$ in the strings
         and their states. Here this would consist of a definition of
         these relations and operations in terms of sums of products of
         qukit A-C operators.  Then one would have to prove that these
         relations and operations satisfy the axioms and theorems for
         the number type being considered.

         In this paper these steps have not been
         provided. Instead the treatment is more like that in
         mathematical analysis textbooks in that descriptions of
         the basic arithmetic relations and
         operations are more directly based on the required axiomatic
         properties and not on algorithms for operations on strings of
         qukits. In essence the treatment here is more like a translation
         of the properties as described in textbooks into the language
         of quantum mechanics for strings of qukits and assuming that
         the relevant proofs apply in this case
         also. A full treatment would require first detailed
         definitions of the basic arithmetic operations and then
         proofs that they satisfy the appropriate axioms. Some
         details are given in \cite{BenRNQM,BenRRCNQT}.

         The invariance of axioms and theorems under
         transformations should be distinguished from the
         covariance of their expressions in various representations.
         Consider for example the axiom $x+0=x$ which says that $0$
         is the additive identity. The expression of this axiom in
         the space $\mathcal{FB}^{X}_{k,m,g}$ for $X=N,I,Ra$ is (again
         $h$ is suppressed)
         \begin{equation}\label{0addid}\begin{array}{l}
         \tilde{+}_{A,k,m,g}|\g,m,s,l,u\rangle_{k,g}|+,m,0\rangle_{k,g}\\
         \hspace{0.5cm}= |\g,m,s,l,u\rangle_{k,g}|+,m,0\rangle_{k,g}
         |\g,m,s,l,u\rangle_{k,g}\end{array}\end{equation}
         for all $\g,s.$ Here $|+,m,0\rangle_{k,g}=\cd_{+,m}(\ad_{k})_{0,m}
         |0\rangle$ is a base $k$ qukit state for the number zero.

         Under the action of $\tilde{T}_{1}$
         this axiom expression becomes\begin{equation}
         \label{Trans0add}\begin{array}{c}\tilde{+}_{A,k,m+1,g}
         |\g,m+1,\sp,l+1,u+1\rangle_{k,g}|+,m+1,0\rangle_{k,g} \\ \hspace{0.5cm}=
         |\g,m+1,\sp,l+1,u+1\rangle_{k,g}|+,m+1,0\rangle_{k,g}\\
         \hspace{1cm}\times|\g,m+1,\sp,l+1,u+1\rangle_{k,g}.
         \end{array}\end{equation}Here $\sp(j+1)=s(j).$

         For $\tilde{W}_{\kp,k}$ the axiom expression is
         \begin{equation}\label{W0add}\begin{array}{l}
         \tilde{+}_{A,\kp,m,g}\tilde{W}_{\kp,k}
         |\g,m,s,l,u\rangle_{k,g}\tilde{W}_{\kp,k}
         |+,m,0\rangle_{k,g}\\ \mbox{}\hspace{0.5cm}=
          \tilde{+}_{A,\kp,m,g}|\g,m,\spp,\lp,\up\rangle_{\kp,g}
         |+,m,0\rangle_{\kp,g}\\ \mbox{}\hspace{1cm}=
         |\g,m,\spp,\lp,\up\rangle_{\kp,g}|+,m,0\rangle_{\kp,g}
         |\g,m,\spp,\lp,\up\rangle_{\kp,g} \end{array}\end{equation}
         This makes sense only for $\kp,k$ for which $\tilde{W}_{\kp,k}$
         is defined. For these the state $|\g,m,\spp,\lp,\up
         \rangle_{\kp,g}$ represents the same number as does
         $|\g,m,s,l,u\rangle_{k,g}.$

         For $U_{k}$ the expression of the axiom is unchanged
         except that it refers to a different basis or reference
         frame. Use of Eq. \ref{Ugg'} gives\begin{equation}
         \label{U0add}\begin{array}{l}
         \tilde{+}_{A,k,m,g'}U_{k}|\g,m,s,l,u\rangle_{k,g}
         U_{k}|+,m,0\rangle_{k,g}\\ \mbox{}\hspace{0.5cm}=
         \tilde{+}_{A,k,m,g'}|\g,m,s,l,u\rangle_{k,g'}|+,m,0\rangle_{k,g'}
         \\ \mbox{}\hspace{1cm}=|\g,m,s,l,u\rangle_{k,g'}|+,m,0\rangle_{k,g'}
         |\g,m,s,l,u\rangle_{k,g'}.\end{array}\end{equation}  Recall
         from Eq. \ref{adU} that the AC operators for the states in
         the changed basis are linear superpositions of the AC operators
         for the states in the original basis.

         The invariance of the axioms and theorems should be
         distinguished from the quantum
         mechanical property of conservation of probability for
         any unitary transformation. For any state $\psi,$ unitary
         transformation $V$ and property expressed by a projection
         operator $P$, one always has $(V\psi,P_{V}V\psi)=(\psi,P\psi)$
         where $P_{V} =VPV^{\dag}.$ However the property expressed
         by $P$ is invariant with respect to $V$ if and only if $P$
         commutes with $V$, $PV=VP.$ It follows that if $P$ expresses
         any of the axioms, then $P$ should commute with
         $\tilde{T}_{1},\tilde{T}_{2},\tilde{W}_{k',k},U_{k}.$

         An example of a property that is not invariant is
         $=_{A,k,m,g}.$ This is used to express arithmetic equality
         in axioms such as $"x+0=x"$.  The problem is that the
         corresponding projection operator, $\tilde{P}_{=_{A,k,m,g}},$ defined by Eq.
         \ref{PeqAkmg} is not invariant under the action of $\tilde{T}_{1},
         \tilde{W}_{k',k},$ or $U_{k}.$  One has\begin{equation}\label{invequal}
         \begin{array}{c}\tilde{T}_{1}\tilde{P}_{=_{A,k,m,g}}=
         \tilde{P}_{=_{A,k,m+1,g}}\tilde{T}_{1} \\ \tilde{W}_{k',k}
         \tilde{P}_{=_{A,k,m,g}} = \tilde{P}_{=_{A,k',m,g}}\tilde{W}_{k',k} \\
         U_{k}\tilde{P}_{=_{A,k,m,g}} = \tilde{P}_{=_{A,k,m,g'}}U_{k}.\end{array}
         \end{equation}

          Part of the invariance lack is due to the fact that $=_{A,k,m,g}$
          and the corresponding projection operator $\tilde{P}_{=_{A,k,m,g}}$
          are locally defined. Invariance with respect to
          $\tilde{T}_{1}$ can be obtained by  an expanded definition of
          arithmetic equality $=_{A,k,g}$ given by $=_{A,k,g}\leftrightarrow
          \exists{m}(=_{A,k,m,g})$ with a corresponding projection operator
          \begin{equation}\label{PAkg}\tilde{P}_{=_{A,k,g}}=\sum_{m}
          \tilde{P}_{=_{A,k,m,g}}.\end{equation}( Eq. \ref{PeqAkmg} with
          the sum over $h,h'$ shows that the operator is invariant under
          $\tilde{T}_{2}.$)

          One would like to complete the process by expanding the
          definitions of arithmetic relations and operations to be
          invariant under $\tilde{W}_{k,k'}$ and $U_{k}.$ However
          there are problems in that $U_{k}$ depends on $k.$  This
          precludes use of a summation over $k$
          because this causes problems with the property that
          the gauge transformations $U_{k}$ depend on $k$.

          To see the problem one notes that although $P_{=_{A,k,g}}$
          is invariant under $\tilde{T}_{1},\tilde{T}_{2},$ it is not
          invariant under either $\tilde{W}_{k',k}$ or $U_{k}.$ The lack
          of $U_{k}$ invariance follows from  Eq. \ref{Pssp}
          which shows that\begin{equation}\label{PUk}\begin{array}{l}
          U_{k}\tilde{P}_{\g,[s],k,h,g}U^{\dag}_{k}=U_{k}\sum_{s'\sim_{0}s}
          \tilde{P}_{|\g,(m,h),s',l,u\rangle_{k,g}}U^{\dag}_{k} \\ \hspace{0.5cm}=
          \sum_{s'\sim_{0}s}\tilde{P}_{U_{k}|\g,(m,h), s',l,u\rangle_{k,g}}
          =\tilde{P}_{\g,[s],k,h,g'}.\end{array}\end{equation} Here the
          $g$ and $g'$ subscripts have been inserted to show the differences
          in the projection operators.

          Two approaches to this problem seem possible. One is based
          on the use of gauge invariant representations of qukit
          states based on irreducible representations of $SU(k)$
          \cite{Byrd,Kempe,Bartlett}.  The other is based on the
          possible use of gauge theory \cite{Gross,Montavy} to
          express the invariance of the arithmetic properties
          by means of an action. Further investigation of this
          problem will be left to future work.

           \section{Composite and Elementary Qukits}\label{CEQ}

           So far the qukit components of strings are considered to
           be different systems for each value of $k.$ A $k$ qukit is
           different from a $k^{\p}$ qukit just as a spin $k$ system
           is different from a spin $k^{\p}$ system. This leads to a
           large number of different qukit types, one for each value
           of $k.$ However, the dependence of the properties of the
           base changing operator $\tilde{W}_{k',k}$ on the prime
           factors of $k$ and $k'$ suggests that one consider qukits
           $q_{k}$ as composites $q_{c_{k}}$ of prime factor qukits
           $q_{p_{n}}$. In general the relation between the
           base $k$ $q_{k}$ and the composite base $k$
           $q_{c_{k}}$ is given by\begin{equation}\label{qck}
           q_{c_{k}}=q_{p_{j_{1}}}^{h_{1}}q_{p_{j_{2}}}^{h_{2}}\cdots
           q_{p_{j_{n}}}^{h_{n}}.\end{equation} where (Eq. \ref{kprfa})
           $$k=p^{h_{1}}_{j_{1}}p^{h_{2}}_{j_{2}}\cdots p^{h_{n}}_{j_{n}}$$.
           Simple examples of this for $k=10$ and $18$ are
           $q_{c_{10}}=q_{2}q_{5}$ and $q_{c_{18}}=q_{2}q_{3}q_{3}.$

           The observation that for each $k$ there is a smallest $k'$
           with the same prime factors and its relevance to the
           properties of $\tilde{W}_{k',k}$ suggest the importance
           of the $q_{c_{k'}}$ where the powers of the prime factors
           are all equal to $1$(Eq. \ref{kpprfa}) \begin{equation}
           \label{qckp} q_{c_{\kp}}=q_{p_{j_{1}}}q_{p_{j_{2}}}\cdots
           q_{p_{j_{n}}}.\end{equation} A particular example of this
           for $k_{n},$ the product of the first $n$ prime numbers, is
           shown by (Eq. \ref{kbase})\begin{equation} \label{qckn}
           q_{_{c_{k_{n}}}}=q_{2}q_{3}q_{5}\cdots q_{p_{n}}
           \end{equation}.

           These considerations suggest a change of emphasis in
           that one should regard  prime number qukits $q_{p_{n}}$
           as basic or elementary and the qukits $q_{k}$ as composites
           of the elementary ones. In this case one would want to consider
           possible physical properties of the elementary qukits and
           how they interact and couple together to form composites.
           This is a subject for future work as the emphasis here is on
           arithmetic properties, not on physical properties.  It
           is, however, intriguing to note that if the prime number
           $q_{p_{n}}$ are considered as spin systems with spin
           $s_{n}$ given by $2s_{n}+1=p_{n}$, then there is
           just one fermion, $q_{2}$.  All the others are bosons.

           As was the case for strings of $q_{k},$ one wants to
           represent numbers by states of finite strings of composite
           $q_{c_{k}}$. In general, this involves replacing the $k$
           dimensional Hilbert space $\mathcal{H}_{k}$ at each site
           in $I\times I$ by a product space\begin{equation}\label{Hck}
          \mathcal{H}_{c_{k}}=\mathcal{H}_{p_{j_{1}}}^{h_{1}}\otimes
          \cdots\otimes\mathcal{H}_{p_{j_{n}}}^{h_{n}}\end{equation}and
          then following the development in the previous sections to
          describe number states. In particular the gauge fixing would
          apply to each component space in Eq. \ref{Hck} for each
          location in $I\times I.$

          In the following, consideration will be limited to the
          simpler case where all the powers $h_{i}=1$ as in Eq.
          \ref{qckp}. In addition the elementary
          $q_{p_{j}}$ in $q_{c_{k'}}$ at each site in $I\times I$
          will be considered noninteracting. In this case the AC
          operators $(\ad_{k'})_{\alpha,(j,h)}\;(a_{k'})_{\alpha,
          (j,h)}$ with $\alpha =0,1,\cdots,k'-1$ for $q_{k'}$ at
          site $(j,h)$ in state $\alpha$ are replaced by products of
           AC operators for the component elementary $q_{p_{j}}.$
           One has\begin{equation}\label{ACqkqck}\begin{array}{l}
          (\ad_{k^{\p}})_{\alpha,(j,h)}=V_{k'}(\ad_{p_{j_{1}}})_{d_{1},(j,h)}
          (\ad_{p_{j_{2}}})_{d_{2},(j,h)}\cdots(\ad_{p_{j_{n}}})_{d_{n},(j,h)}
          \\ (a_{k^{\p}})_{\alpha,(j,h)}=(a_{p_{j_{1}}})_{d_{1},(j,h)}
          (a_{p_{j_{2}}})_{d_{2},(j,h)}\cdots(a_{p_{j_{n}}})_{d_{n},(j,h)}
          V^{\dag}_{k'}\end{array} \end{equation}where for each $i=1,\cdots,n$
          $d_{i}$ is an element of $0,1,\cdots,p_{j_{i}}-1$ and
          \begin{equation}\label{alphabeta}\alpha =\beta(d_{1},\cdots,d_{n}).
          \end{equation}

          Here $\beta:\{\{0,\cdots,p_{j_{1}}-1\}\times\cdots
          \times\{0,\cdots,p_{j_{n}}-1\}\}\rightarrow
          \{0,\cdots,k'-1\}$ maps states of the $n-tuple$ of prime
          number qukits of $q_{c_{k'}}$ to states of $q_{k'}.$
          $V_{k'}$ is a unitary operator that maps states in
          $\mathcal{H}_{p_{j_{1}}}\times\cdots \mathcal{H}_{p_{n}}$
          to states in $\mathcal{H}_{k'}.$ For any $q_{c_{k}}$ in
          state $|d_{1},\cdots,d_{n}\rangle,$
          $V_{k}|d_{1},\cdots,d_{n}\rangle =|\alpha\rangle.$

          The replacements given above can be used for states of strings of
          composite $q_{c_{k^{\p}}}$ systems. Each state $|\g,(m,h),
          \sp,l,u\rangle_{c_{k'}}$ is given by strings of AC
           operators as\begin{equation}\label{cmpstr}\begin{array}{l}
           |\g,(m,h),\sp,l,u\rangle_{c_{k'}}
          =\cd_{\g,(m,h)}\{(\ad_{p_{j_{1}}})_{d'_{1}(u,h),(u,h)}\cdots
          (\ad_{p_{j_{n}}})_{d'_{n}(u,h),(u,h)}\} \\ \cdots\{(\ad_{p_{j_{1}}}
          )_{d'_{1}(l,h),(l,h)} \cdots(\ad_{p_{j_{n}}})_{d'_{n}(l,h),(l,h)}\}
          |0\rangle.\end{array}\end{equation} Here $s'$ is a function from
          $I\times I$ to $\{\{0,\cdots,p_{j_{1}}\}\times\cdots
          \times\{0,\cdots,p_{j_{n}}-1\}\}$ and $d_{i}(j,h)$ is the
          $ith$ component of $s'(j,h).$

          The requirement that states of the form
          $|\g,(m,h),\sp,l,u\rangle_{c_{k'},g}$ represent numbers is
          based on an ordering of the basis states of $q_{c_{k}}$,
          or, what is equivalent, an ordering of the  $n-tuples$ in
          the range set of $s'$.   The definitions of
          arithmetic relations and operations for these states must
          respect the ordering and they must satisfy the relevant axioms
          and theorems for the type of number being considered.

          States of strings of composite qukits  and their arithmetic
          properties can be directly related to states of strings of
          $q_{k'}$ where  $k' =p_{j_{1}}\cdots p_{j_{n}},$ Eq. \ref{kpprfa}.
          The arithmetic properties of the states $|\g,(m,h),s,l,u\rangle_{k',g},$
          where $s$ is a function from $I\times I$ with values in
          $\{0,1,\cdots,k'-1\},$ and the ordering of the states is based
          on the map $\b,$ Eq. \ref{alphabeta} which is required to be
          order preserving.

          Also Eqs. \ref{ACqkqck} and \ref{cmpstr} and the
          unitarity of $V_{k'}$ should give the result that
          $|\g,(m,h),s,l,u\rangle_{k'}$ and $|\g,(m,h),s',l,u
          \rangle_{c_{k'}}$ represent the same number even
          though they are very different quantum mechanically.
          This is a nontrivial requirement. It depends on
          $V_{k'},$ mappings of the ordering, arithmetic
          relations and operations on states of $q_{c_{k'}}$
          strings to those for states of $q_{k'}$ strings, and
          proof of the invariance of the relevant axiomatic and
          theorem properties under the action of $V_{k'}.$

          The description of the transformation operations
          $\tilde{T}_{1},\tilde{T}_{2},\tilde{W}_{k',k},U_{k'}$ can
          be extended to apply to the composite qukit strings.
          $\tilde{T}_{1}$ and $\tilde{T}_{2}$ shift all elementary
          $q_{p_{j}}$ in $q_{c_{k'}}$ by one unit in either
          direction. The base changing operator
          $\tilde{W}_{c_{k'},c_{k}}$ changes states of $q_{c_{k}}$
          strings to states of $q_{c_{k'}}$ strings that should
          represent the same number.  Note that the expression of
          $\tilde{W}_{c_{k'},c_{k}}$ in terms of sums of products of
          AC operators will include the annihilation of many
          component elementary qukits in $q_{c_{k}}$ and creation
          of many that are components of $q_{c_{k'}}.$

          The description of gauge transformations $U_{c_{k}}$
          applied to states of $q_{c_{k}}$ is interesting. If
          $q_{c_{k}}$ is composed of elementary $q_{p_{j}}$ as given
          by Eq. \ref{qck}, then $U_{c_{k}}$ is a map from $I\times
          I$ to elements of $U(p_{j_{1}})^{h_{1}}\times\cdots\times
          U(p_{j_{n}})^{h_{n}}.$ Here $U(p_{j_{i}})$ is the unitary group
          of prime dimension $p_{j_{i}}.$ For the special cases of
          Eq. \ref{qckp} and \ref{qckn} $U_{c_{k'}}$ and
          $U_{c_{k_{n}}}$ take values in $U(p_{j_{1}})
          \times\cdots\times U(p_{j_{n}})$ and in
          $U(p_{1})\times\cdots\times U(p_{n})$ respectively. Since
          $U(p_{j}) =U(1)\times SU(p_{j})$ the values of
          $U_{c_{k_{n}}}$ can be represented as elements of
          \begin{equation}\label{SUprod}\begin{array}{l}
          U(1)\times SU(p_{1})\times SU(p_{2})\times\cdots\times
          SU(p_{n})\\ \mbox{}\hspace{1cm}=U(1)\times SU(2)\times SU(3)\times
          SU(5)\times\cdots\times SU(p_{n}).\end{array}\end{equation} Here the
          phase factor elements in $U(1)$ for each elementary qukit
          have been combined into one phase factor for the composite
          $q_{c_{k_{n}}}.$

          The discussion so far suggests that, as far as quantum
          theory representations of natural numbers, integers, and
          rational numbers are concerned, it is sufficient to limit
          components of gauge transformations to  products of elements
          of $U(1)$ and products of elements of $SU(p)$ groups
          where $p$ is a prime number. Furthermore it is sufficient
          that, for each prime $p,$ elements of $SU(p)$ occur at most
          once in the product. It is also sufficient to limit
          components to products of the form of Eq. \ref{SUprod} for
          $n=1,2,\cdots$ as these will include representations for
          all rational numbers.

         \section{Discussion}
         There are several additional aspects of the work presented
         here that would benefit from further discussion.  The
         approach in which one regards $q_{k}$ qukits for arbitrary
         $k$ as composites $q_{c_{k}}$ of elementary prime number
         qukits $q_{p}$ where $p$ is a prime number has an
         interesting property.  To see this one  recalls that the
         domain and range of the base changing operator
         $\tilde{W}_{k',k},$ as a map from
         $\mathcal{FB}^{Ra}_{k,g}$ to $\mathcal{FB}^{Ra}_{k',g},$
         depend on the prime factors of $k$ and $k'.$ It follows
         that one must be able to determine the prime factors of $k$
         and $k'$ to obtain the properties of this operator.

         However, as is well known, there is no known classical
         algorithm for efficiently obtaining the prime factors of an
         arbitrary large number. The only known efficient algorithm
         for factorization \cite{Shor} is quantum mechanical and is
         based on the exponential speedup possible for quantum
         computers.

         Working with composite $q_{c_{k}}$ systems avoids the
         factorization problem completely. In this case one works at
         the outset with composites containing different numbers of
         prime components. Determination of the value of $k$  for a
         given set of prime components is efficient and
         straightforward. It is not clear if the value of $k$
         is even needed with this approach. The possible exception is
         its use as a label or coding, as in $\tilde{W}_{k',k}$,
         for arbitrary sets of prime components $q_{p}.$

         The observation that the only known efficient factoring
         algorithm is quantum mechanical provides some support for
         the emphasis in this paper on quantum representations of
         numbers.  The possibility of a deeper connection between
         quantum computation algorithms and the spaces of quantum
         representations of numbers presented here is a subject
         for future work.

         It is useful to reemphasize the importance of the requirement
         that the action of the transformation operators,
         $\tilde{T}_{1},\tilde{T}_{2},\tilde{W}_{k',k},U_{k}$ on
         basis states $|\g,(m,h),s,l,u\rangle_{k,g}$ preserve
         the number representing property of the states. For
         instance, \begin{equation}\label{sameno}\begin{array}{c}
         \tilde{W}_{k',k}|\g,(m,h),s,l,u\rangle_{k,g}
         =|\g,(m,h),s'.l'.u'\rangle_{k',g} \\ \mbox{and }
         U_{k}|\g,(m,h),s,l,u\rangle_{k,g}=|\g,(m,h),s,l,u\rangle_{k,g'}
         \end{array}\end{equation} must represent the same number as
         does $|\g,(m,h),s,l,u\rangle_{k,g}.$ Note that the state
         $|\g,(m,h),s,l,u\rangle_{k,g'}$ is different from
         $|\g,(m,h),s,l,u\rangle_{k,g}$ in that it is described
         using transformed AC operators given in Eqs. \ref{Urast}
         and \ref{adU}.

         Verification of this requirement involves showing that
         the basic arithmetic operations and relations must have
         the properties described by the relevant axioms and theorems.
         As noted, this is not obvious because the definitions of the
         operations and relations are based on algorithms for basic
         quantum operations on states of $q_{k}$ or $q_{c_{k}}$ in
         strings. Here this problem was bypassed by simply assuming that
         the arithmetic relations and operations have the requisite
         properties. Some work in this direction is given in
         \cite{BenRNQM,BenRRCNQT}.

         It is hoped to examine this in more detail in the future.
         One possible approach is to follow the development of gauge
         theories \cite{Mack,Montavy} and define parallel transport
         for the property "the same number as". The problem here is
         that one does not have  an action or Lagrangian for the
         axioms of a theory. Recall that the axioms and theorems of
         a theory are invariant under the transformations described
         here.  If such an action could be discovered then its
         invariance would  give the desired result.

         It is amusing to note that for strings of composite $q_{c_{k_{n}}}$
         where $q_{c_{k_{n}}}=q_{2}q_{3}q_{5}\cdots q_{p_{n}},$ Eq.
         \ref{qckn}, gauge transformations have the form $U(1)\times
         SU(2)\times\cdots\times SU(p_{n})$, Eq. \ref{SUprod}.
         Invariance of the axioms and theorems under these
         transformations reminds one of the standard model of
         physics \cite{Cott,Novaes,Roy} which requires invariance of
         a Lagrangian under the gauge transformation given above
         for $n=3.$  Whether there is any further connection or not
         remains to be seen.

         The emphasis in this paper has been on the quantum
         representations of numbers in $N,I$ and $Ra$ as
         mathematical systems. The question arises regarding  the
         connection of these results to physics.  Making such a
         connection is essential if one is to make headway towards a
         coherent theory of mathematics and physics together. Here
         an indirect connection is implied by using physical
         concepts of representations parameterized by points in a
         parameter set and induced transformations of the
         representations. The invariance of relevant axioms and
         theorems under the transformations and the decomposition of
         $q_{k}$ into composites  $q_{c_{k}}$ of elementary $q_{p}$
         are also concepts used in physics. Also, limitation of
         the representations to  single strings of qukits is  dictated
         by physical considerations.

         These connections are all  indirect.  A more direct
         one may be to take advantage of the fact that, as units of
         quantum information, qukits can represent both physical and
         mathematical systems. This fact is used extensively in work
         on quantum computation \cite{Nielsen} where states of qubit
         strings represent both numbers in quantum computers and states
         of physical systems with associated dynamics that are models
         of  quantum computers (\cite{Vincenzo} and references cited
         therein). Future work will tell if this approach is useful.

         \section{Summary}
         In this paper spaces of quantum representations of natural
         numbers, integers, and rational numbers were described. The
         space is based on a parameterization of the different
         representations, as states of finite strings of qukits,
         by  $4-tuples$ $k,(m,h),g$ which are elements of a
         parameter  set $N\geq 2\times I\times I\times \{g\}.$
         Here $(m,h)$ locates the string in a $2$ dimensional
         integer lattice $I\times I$, $k\geq 2$ is the qukit base,
         and each $g$ is a gauge fixing function that chooses the
         qukit basis in  the $k$ dimensional Hilbert space of states
         at each lattice point.

         Associated with each point $k,(m,h),g$ is a representation
         space $\mathcal{FB}^{X}_{k,(m,h),g}$ consisting of a
         Fock space $\mathcal{F}^{X}_{k,(m,h)}$ of states of finite
         strings of base $k$ qukits at locations $(j,h)$ for all
         $j\epsilon I$, and a set,
         $\mathcal{B}_{k,(m,h),g}=B_{2,m,h}\cup
         \{B_{k,j,h,g}:j\epsilon I\},$ of basis states. For each $j$
         $B_{k,j,h,g}$ is a basis for $\mathcal{H}^{k}_{j,h}$ and $B_{2,m,h}$
         is the basis for the sign qubit at $(m,h).$ The states
         $|\g, (m,h),s,l,u\rangle_{k,g}$ in $\mathcal{B}_{k,(m,h),g}$
         can be said to span $\mathcal{F}^{X}_{k,(m,h)}.$  These
         states are described by strings of AC operators acting on
         the qukit vacuum state $|0\rangle.$  For $X=N,I,Ra$ the states
         represent natural numbers, integers, or rational numbers
         respectively.  Here $\g = +,-$ denotes the sign, $(m,h)$
         the location of the $"k-al"$ point, and $s$ is a
         $0,1,\cdots,k-1$ valued function on the lattice domain
         $[(l,h),(u,h)].$ Also $l\leq m\leq u.$

         Transformations on the parameter set induce
         transformations on the representation space. The
         properties of the base changing operator,
         $\tilde{W}_{k',k},$ were seen to depend on the relations
         between the prime factors of $k$ and $k'.$ Global and
         local gauge transformations $U_{k}$ were defined as
         functions from $I\times I$ to $U(1)\times SU(k).$ These $k$
         dependent transformations change the gauge or basis choice
         for the Hilbert spaces on $I\times I$.

         The base $k$ was extended to $k=1$ by considering unary
         representations of numbers. It was noted that these are
         unique as they are the only ones that are extensive
         and not representational. The extensivity is seen by
         noting that any collection of systems, e.g. strings of
         $q_{k}$, is an unary representation of a number that is the
         number of systems in the collection. This ties in with the
         ubiquitous presence of $U(1)$ in gauge transformations for
         all $k\geq 2.$ Also the observation that noninteger
         rational numbers do not have an unary representation by
         a \emph{single} string or a \emph{single} collection of
         systems, was seen to tie in neatly with the properties of
         $\tilde{W}_{k',k}.$

         The invariance of arithmetic properties of the qukit string
         under the action of the transformation operators was
         discussed.  In particular the properties expressed by the
         axioms and theorems for each type, $N,I,Ra$ of numbers are
         invariant. It was noted that, the fact that gauge transformation
         operators $U_{k}$ depend on the base, causes problems for
         implementation of the invariance for the base changing
         operators $\tilde{W}_{k,k'}$ and gauge transformations. Two
         possible approaches to the problem, the use of gauge invariant
         representations of qukit states and the use of gauge
         theory, were suggested. However, further work on the problem
         was left to future work.

         It was noted that the dependence of the properties of
         $\tilde{W}_{k',k}$  on the prime number factors of
         $k$ and $k'$, suggests a change in emphasis from qukits as
         $q_{k}$ to composites $q_{c_{k}}$ of elementary prime number
         qukits $q_{p}.$ If $k=(p_{j_{1}})^{h_{1}}\cdots(p_{j_{n}})^
         {h_{n}},$ then $q_{c_{k}}=(q_{p_{j_{1}}})^{h_{1}}\cdots
         (q_{p_{j_{n}}})^{h_{n}}.$ Here the component elementary $q_{p}$
         in a composites were assumed to be noninteracting. Then
         states of each $q_{c_{k}}$ can be represented as $m-tuples$ of
         states of the component $q_{p}$ where $m$ is the number
         of components in $q_{c_{k}}.$  The special case where
         $q_{c_{k_{n}}}=q_{2}q_{3}q_{5}\cdots q_{p_{n}}$ primes was
         described, particularly with respect to gauge
         transformations of the form $U(1)\times
         SU(2)\times\cdots\times SU(p_{n}).$

         Finally, several different aspects of the work presented
         here were discussed.  The emphasis was on the many
         different avenues for future work. In any case it is hoped
         that the ideas presented here will show that studies of
         quantum representations of numbers as states of single strings
         of qukits may be useful and relevant to physics. Numbers \emph{are}
         the foundations of all physical theories. One may also hope
         that this work is a useful small step towards a coherent
         theory of mathematics and physics together.

          \section*{Acknowledgements}
          This work was supported by the U.S. Department of Energy,
          Office of Nuclear Physics, under Contract No.
          DE-AC02-06CH11357.


\begin{thebibliography}{99}
          \bibitem{Wigner}
          E. Wigner, Communications in Pure and Applied Mathematics,
          \textbf{13}, 1-14, (1960).

          \bibitem{Hamming}
          R. W. Hamming, Amer. Mathematical Monthly, \textbf{87},No 2,
          February, (1980).

          \bibitem{Fefferman}
          S. Fefferman, \emph{In the Light of Logic}, Oxford
          University Press, New York, 1998.

          \bibitem{Tegmark1}
            M. Tegmark, Ann. Phys. \textbf{270}, 1 (1998) (Arxiv
            preprint gr-qc/9704009).

            \bibitem{Tegmark2}
            M. Tegmark, Arxiv preprint 0704.0646v1 [gr-qc].

            \bibitem{Schmidhuber}
            J. Schmidhuber, Arxiv preprint quant-ph/0011122v2.

            \bibitem{Weinberg}
            S. Weinberg, \emph{Dreams of a Final Theory},
            Vintage Books, New York, 1994


          \bibitem{Fraenkel}
          A. Fraenkel, Y. Bar-Hillel, A. Levy, \emph{Foundations of Set
          Theory} 2nd Revised Edition, Studies in Logic and the
          Foundations of Mathematics, Vol. 73, North Holland
          Publishing Co. London, 1973.

          \bibitem{Davies}
          P. C. W. Davies, Arxiv preprint, quant-ph/0703041.

          \bibitem{BenTCTPM}
          P. Benioff,  Foundations of Physics,
          \textbf{35}, 1825-1856, (2005).

           \bibitem{Litvinov}
            G. L. Litvinov, V. P. Maslov, and G. B. Shpiz,
            Archives preprint, quant-ph/9904025, v5, 2002.


            \bibitem{Corbett}
            J. V. Corbett and T. Durt, Archives preprint,
            quant-ph/0211180 v1 2002.

            \bibitem{Tokuo}
            K. Tokuo, Int. Jour. Theoretical Phys.,
            \textbf{43}, 2461-2481, 2004.

            \bibitem{Finkelstein}
            D. Finkelstein, \emph{Quantum Relativity},
            Springer-Verlag, Heidelberg (1996)


            \bibitem{Takeuti}
          Gaisi Takeuti, \emph{Two Applications of Logic to
          Mathematics} Kano Memorial Lecture 3, Princeton University
          Press, New Jersey, 1978;  \emph{Quantum set theory}, in:
          E. G. Beltrametti, B. C. van Fraassen, Eds.,
           \emph{Current issues in quantum logic},
           Plenum, pp. 303-322, New York 1981.

          \bibitem{Davis}
          Martin Davis, Internat. Jour. Theoret. Phys.
          \textbf{16},867-874,(1977).

          \bibitem{Schlesinger}
           Karl-Georg Schlesinger, Journal of Mathematical Physics,
           \textbf{40}, 1344-1358 (1999).

           \bibitem{Titani}
           S. Titani and H. Kozawa, Internat. Jour. Theoret. Phys.
           42, 2575-2602, (2003).

           \bibitem{Krol}
          Jerzey Krol, "A Model of Spacetime. The Role of
          Interpretations in Some Grothendieck Topoi", preprint,
          (2006).

        \bibitem{BenRNQM}
        P. Benioff, Algorithmica, \textbf{34}, 529-559, (2002),
        (quant-ph/0103078).

        \bibitem{BenRRCNQT}
        P. Benioff, Arxiv preprint, quant-ph/0508219.

          \bibitem{Nielsen}
          M. A. Nielsen and I. L. Chuang, \emph{Quantum Information
          and Quantum Computation}, Cambridge University Press,
          New York, 2000.

          \bibitem{BenRCRNQM}
          P. Benioff, Phys. Rev. A \textbf{72}, 032314, (2005),
          (quant-ph/0503154).

          \bibitem{Mack}
          G. Mack, Fortschritte der Physik,
          \textbf{29},135-185,(1981).

          \bibitem{Halmos}
          P.Halmos, \emph{A Hilbert Space Problem Book, 2nd
          Edition}, Springer-Verlag, New York, 1982.

          \bibitem{Crypt}
          C. H. Bennett, G. Brassard, and A.K. Ekert,
          "Quantum cryptography", Scientific American,
           pp. 50 - 57, October 1992; C. H. Bennett,
          Phys. Rev. Letters, \textbf{68}, 3121 - 2124, (1992);
          G. Brassard, C. Crépeau, R. Jozsa,  and D. Langlois,
          \emph{A quantum bit commitment scheme provably unbreakable
          by both parties}, Proceedings, 34th Annual IEEE Symposium
          on Foundations of Computer Science, November 1993, pp. 362 - 371.

          \bibitem{Enk}
          S. J. van Enk, Phys.Rev. A \textbf{73},  042306, (2006).

          \bibitem{Byrd}
          Mark S. Byrd, Daniel Lidar, Lian-Ao Wu, and Paolo Zanardi,
          Phys. Rev A \textbf{71}, 052301 (2005).

          \bibitem{Kempe}
          J. Kempe, D. Bacon, D. A. Lidar, and K. B. Whaley, Phys.
          Rev. A \textbf{63}, 042307 (2001).

          \bibitem{Bartlett}
          S. D. Bartlett, T. Rudolph, and R. W. Spekkens, Revs.
          Modern Phys., \textbf{79}, 555-609, (2007).

          \bibitem{Gross}
          D. Gross, \emph{Gauge Theory-Past, Present, and Future}
          Chinese Journal of Physics, 30, 955-972, (1992)
          (Wiki:-gauge theory ref.)

          \bibitem{Montavy}
        I. Montavy and G. M\"{u}nster, \emph{Quantum Fields on a
        Lattice}, Cambridge University Press, New York, NY, 1994.


        \bibitem{Shor}
        P. Shor, in {\it Proceedings, 35th Annual Symposium on the
        Foundations of Computer Science}, S. Goldwasser (Ed), IEEE
        Computer Society Press, Los Alamitos, CA, 1994, pp 124-134; SIAM
        J. Computing, {\bf 26}, 1484-1510 (1997).

          \bibitem{Cott}
          A. N. Cottingham and D. A. Greenwood, \emph{An Introduction
          to the Standard Model of Physics}, Cambridge University
          Press, Cambridge, UK, 1998.

          \bibitem{Novaes}
          S. F. Novaes, Arxiv preprint hep-ph/0001283.

          \bibitem{Roy}
          D. P. Roy, Arxiv preprint hep-ph/9912523.

          \bibitem{Vincenzo}
          D. P. DiVincenzo, Arxiv preprint quant-ph/0002073.

          \end{thebibliography}
           \end{document}